\documentclass[a4paper,12pt]{article}
\pdfoutput=1

\usepackage{jheppub}
\usepackage[utf8]{inputenc}
\usepackage[english]{babel}

\usepackage{bm}
\usepackage{amsfonts}
\usepackage{mathbbol}
\usepackage{amsthm}
\usepackage{amssymb}
\usepackage{mathrsfs}
\usepackage{latexsym}
\usepackage{commath} %\abs is defined here
\usepackage{float}
\usepackage{subfig}
\usepackage{subcaption}

\graphicspath{{images/}}

\newcommand{\be}{\begin{equation}}
\newcommand{\ee}{\end{equation}}
\newcommand{\eps}{\varepsilon}
\newcommand{\HyperF}[4]{{}_{2}F_{1}\left(#1,#2;#3;#4\right)}
\newcommand{\HyperFTransformed}[4]{{}_{2}\tilde{F}_{1}\left(#1,#2;#3;#4\right)}
\newcommand{\const}{\mathbb{c}}
\newcommand{\om}{\omega}
\newcommand{\cut}{_\text{cut}}
\DeclareMathOperator\arctanh{arctanh}
\newcommand{\skipline}{\vspace{\baselineskip}}
\DeclareMathOperator\re{Re}
\DeclareMathOperator\im{Im}
\newlength{\twopanelfigurewidth}

\begin{document}

\title{Spectral form factors for curved spacetimes with horizon}

\author[a,b]{Dmitry S. Ageev,}
\author[a]{Vasilii V. Pushkarev}
\author[a]{and Anastasia N. Zueva}

\affiliation[a]{Steklov Mathematical Institute of Russian Academy of Sciences,\\
Gubkin str. 8, 119991 Moscow, Russia}
\affiliation[b]{Institute for Theoretical and Mathematical Physics, Lomonosov Moscow State University, 119991 Moscow, Russia}

\emailAdd{ageev@mi-ras.ru}
\emailAdd{pushkarev@mi-ras.ru}
\emailAdd{zueva@mi-ras.ru}

\abstract{The spectral form factor is believed to exhibit a special type of behavior called “dip-ramp-plateau” in chaotic quantum systems that originates from random matrix theory. This suggests that the shape of the spectral form factor could serve as an indicator of chaos in various quantum systems. It has been shown recently that the dip-ramp-plateau structure appears in the spectral form factor when the normal modes of a massless scalar field theory in the brick-wall model of the BTZ black hole are treated as eigenvalues of a quantum Hamiltonian. At the same time, the level spacing distribution of these normal modes differs from that associated with random matrix theory ensembles. In this paper, we extend the results for BTZ background to the case of non-zero mass of the field, study the generalized spectral form-factor, and consider the same context for another non-trivial background --- de Sitter space. We compare the generalized spectral form factor for simple integrable quantum systems and for backgrounds with a horizon to the behavior predicted by random matrix theory. As a result, we confirm that BTZ and de Sitter brick-wall models are highly distinct integrable systems that exhibit the dip-ramp-plateau structure of the SFF but differ in the structure of the three-level generalized spectral form factor from the one predicted by random matrix theory. This raises the question on whether the DRP structure is an indicator of what is known as quantum chaos.
}

\maketitle

%---------------%
\section{Introduction}

Quantum gravity provides different opportunities to study unusual features of quantum systems inaccessible in flat space~\cite{Bousso:2022ntt}. The fine-grained properties of quantum information, holographic correspondence, and strongly coupled phenomena are just a few of the research topics in the context of black holes. Another important examples are various manifestations of chaos~\cite{Sekino:2008he, Shenker:2013pqa, Maldacena:2015waa}. Classical chaos, being a relatively well-understood phenomenon, becomes an elusive concept, when  extended to the quantum realm~\cite{Haake:2010fgh}. An archetypal model used for comparison with quantum systems is random matrix theory (RMT). Within the RMT paradigm, there are quantum mechanical quantities to be studied to determine whether the system is chaotic or not. These quantities include eigenvalue fluctuation correlations~\cite{Bohigas84}, such as the nearest-neighbor level spacing distribution, with a vast number of analytical results available for large matrices drawn from random ensembles. 

In quantum field theory, it is not obvious how to compare its spectral properties with those of random matrices, which are finite-dimensional systems with a discrete spectrum. A step towards understanding this problem was taken in~\cite{Rosenhaus:2020tmv}, where it was conjectured that chaotic behavior in quantum field theory can be detected by analyzing the S-matrix. Indeed, in~\cite{Savic:2024ock}, by studying highly excited string scattering amplitudes, the spacing distribution between the phases of the complex eigenvalues of the S-matrix was shown to indicate chaotic dynamics. In~\cite{Bianchi:2022mhs, Bianchi:2023uby}, it was proposed that to quantify this behavior one can estimate the distribution of distances between successive extrema of an amplitude. A similar idea was previously applied to Riemann zeta function in~\cite{berry2005riemann, odlyzko1987distribution}, where it was observed that the corresponding distribution, despite being completely deterministic, closely resembles that of random matrix ensembles.

One can accumulate the spectrum into a single quantity called the spectral form factor (SFF). It was introduced as the Fourier transform of the two-level correlation function of level spacings~\cite{Haake:2010fgh, Liu:2018hlr} but in the context of strongly coupled systems, it can be essentially calculated by ensemble averaging of the evolution operator or the analytically continued partition function~\cite{Cotler:2016fpe}. This provides another possible way to deal with quantum field theories. Random matrix theories are believed to be characterized by the specific behavior of the spectral form factor called the dip-ramp-plateau (DRP) structure by the featuring details of time evolution~\cite{Cotler:2016fpe}. The ramp part, which is a period of linear growth on a log-log scale at intermediate times, is associated with level repulsion, while the absence of late-time decay indicates the discreteness of the spectrum. Comparison between the structure of the SFF obtained for some quantum system and the random matrix DRP-structure may reveal whether the dynamics of the system is chaotic. This is similar to comparison between the probability density functions of nearest-neighbor level spacing distributions.

It was recently discovered in~\cite{Das:2022evy}, that the spectrum of the normal modes of massless scalar field on the BTZ black hole background with a stretched horizon (BTZ brick-wall model) produces an RMT-like spectral form factor even without additional ensemble averaging. In this approach, normal modes are interpreted as energy eigenvalues of corresponding quantum-mechanical system (see~\cite{Das:2023ulz, Das:2023xjr, Krishnan:2023jqn, Banerjee:2024dpl, Jeong:2024oao, Banerjee:2024ivh} for further development of this model).

In the initial version of the setup~\cite{Das:2022evy}, despite the SFF having the same properties (DRP structure) as in RMT, the level spacing distribution does not resemble the RMT behavior. However, as it was shown in~\cite{Das:2023ulz}, the level repulsion can also be reproduced if one introduces an angular-dependent boundary condition at the stretched horizon with the corresponding phases being distributed according to a Gaussian distribution. A detailed analysis of spectral statistics for the Gaussian-distributed boundary conditions both for bosonic and fermionic fields was performed in~\cite{Jeong:2024oao}. This randomness naturally leads to spectral statistics similar to RMT.

It is curious that the level spacing distribution without any additional assumptions does not resemble the statistics of random matrix theory, while the SFF does. Furthermore, the ramp part is not observed for empty AdS spacetime~\cite{Das:2022evy}, indicating that the horizon is required for the DRP structure to emerge. We explore this model in different directions and also study de Sitter spacetime, which features a horizon and positive curvature, raising the question of whether the fact that the curvature is negative is crucial for the described picture.

We start with a mild generalization of the results obtained earlier in~\cite{Das:2022evy} for massless fields on the BTZ black hole background. We find that for the massive field for small number of angular momenta modes one cannot reproduce the DRP structure. Adding even a relatively small mass requires taking into account a significantly larger number of angular momenta to reproduce the ramp part.

After that, we consider the normal modes of a scalar field in the static patch of three-dimensional de Sitter spacetime and calculate their SFF. We find the similar structure of the level spacing distribution, as well as the behavior of the spectral form factor as for the BTZ black hole, thus demonstrating that the sign of the curvature does not influence the observations.

Then, we extended the calculation of the SFF to a recently proposed the generalized spectral form factor (gSFF)~\cite{Wei:2024ujf}. It can be considered a refined version of the SFF such that its definition allows accounting for correlations between any arbitrary number of levels. In contrast to the SFF, its generalized version is a complex-valued quantity. Considering the curious situation in which we have the level spacing distribution that does not coincide with any of the Gaussian ensemble distributions, but a DRP-type spectral form factor, it could help to clarify whether this refined quantity still resembles the behavior characteristic of random matrix theory. We compare the specific behavior of regular systems such as the harmonic oscillator and the rectangular billiard to the RMT results. The two- and three-level gSFF confirm the similarity observed in the context of standard SFF for the BTZ black hole and de Sitter but the late-time behavior of the three-level one differs from that of both regular systems and RMT.

Let us highlight our results.
\begin{itemize}
    \item The ordinary spectral form factor and its generalizations, as well as the level spacing distribution exhibit a similar structure for the BTZ and de Sitter normal modes, despite the opposite signs of the curvature and horizon specifics of the spacetimes.
    \item The DRP structure of the ordinary spectral form factor is less distinct in the massive case for both spacetimes and reveals only for higher cutoff values, i.e., a significantly larger number of normal modes should be included for it to appear.
    \item The real part of the two-level generalized spectral form factor reveals consistency with the ordinary SFF behavior and the complex part has a form similar to that obtained for an ensemble of random matrices.
    \item The three-level gSFF has a shape that differs from both RMT and simple integrable models. The behavior of the real part is highly oscillatory, but only at late times these oscillations are close to those of simple integrable models. The behavior of the imaginary part matches that of random matrices, but differs from simple integrable systems at intermediate times.
\end{itemize}

The structure of the paper is as follows. In Section~\ref{sec:BTZ}, we study BTZ black hole normal modes for massive scalar field and the corresponding SFF. In Section~\ref{sec:dS}, we repeat our calculations for de Sitter spacetime. In Section~\ref{sec:GSFF}, we introduce generalized spectral form factors, study them for BTZ black hole and de Sitter as well as for simple integrable models. The final section is devoted to conclusions.

%--------------------------%
\section{BTZ black hole: massive fields\label{sec:BTZ}}
\subsection{Scalar field on BTZ black hole background}

As a warm-up, let us review in detail a massive scalar field normal modes on the BTZ black hole background first considered in~\cite{Ichinose:1994rg}. The BTZ spacetime is given by a metric of the following form,
\be
    ds^2 = -\frac{r^2 - r_h^2}{\ell^2}dt^2 + \frac{\ell^2}{r^2 - r_h^2}dr^2 + r^2d\varphi^2, \quad r > r_h,
    \label{eq:BTZ}
\ee
having a horizon at ${r = r_h}$ and $\varphi$~is assumed to be a compact coordinate, $\ell$~is the AdS radius.

We study a massive scalar field minimally coupled to this background metric, which satisfies the equation of motion
\be\label{eq:EoM}
    (\Box - m^2)\Phi = 0, \quad \text{where} \quad \Box\Phi \equiv \frac{1}{\sqrt{-g}}\partial_\mu(\sqrt{-g}g^{\mu\nu}\partial_\nu\Phi).
\ee
Separation of variables provides an ansatz
\be\label{eq:ansatz}
    \Phi(t, r, \varphi) = \sum_{\om,\,J}e^{-i\om t}e^{iJ\varphi}\frac{\phi_{\om,\,J}(r)}{\sqrt{r}},
\ee
which, when substituted into~\eqref{eq:EoM}, yields a radial equation defining $\phi_{\om,\,J}(r)$, where $J$~is an integer. In further calculations, we omit the subscripts~$\om$ and~$J$. For the metric~\eqref{eq:BTZ}, this equation has the form
\be\label{eq:radial}
    (r^2 - r_h^2)^2\frac{d^2\phi(r)}{dr^2} + 2r(r^2 - r_h^2)\frac{d\phi(r)}{dr} + \ell^4\om^2\phi(r) - V_\text{eff}(r)\phi(r) = 0,
\ee
where the effective potential is given by
\be
    V_\text{eff}(r) = (r^2 - r_h^2)\left[\frac{1}{r^2}\left(\ell^2 J^2 + \frac{r_h^2}{4}\right) + \ell^2 m^2 + \frac{3}{4}\right].
\ee

The radial equation admits two independent solutions; imposing Dirichlet conditions outside the horizon yields the normal modes. At the stretched horizon located at some finite coordinate ${r_0 > r_h}$, we define
\be
    \phi(r_0) = 0,
\ee
while as a second boundary condition, we require the solution to vanish as ${r \to \infty}$,
\be
    \lim\limits_{r\,\to\,\infty} \phi(r) = 0.
    \label{eq:outer_bc}
\ee
Since the effective potential grows as~$r^2$ at large~$r$, such conditions determine a discrete set of real values of~$\om$ known as the normal modes considered typically in brick-wall models~\cite{tHooft:1984kcu}. In contrast, quasi-normal modes, which are complex-valued, arise for the infalling boundary condition at the horizon~\cite{Keski-Vakkuri:1998gmz}. Each normal mode in our case depends on two integer numbers~$n$ and~$J$, which label the radial and angular modes, respectively. In numerical calculation, one needs to define cutoffs~$n\cut$ and~$J\cut$, which determine how many modes are taken into account.

Substitution of the form
\be
    \phi(r) \to \left(1 - \frac{r_h^2}{r^2}\right)^\alpha \left(\frac{r_h}{r}\right)^\beta \phi\left(\frac{r_h^2}{r^2}\right),
\ee
with 
\be
    \alpha = \frac{i\ell^2\om}{2r_h}, \quad \beta = \frac{1}{2} + \nu, \quad \nu = \sqrt{1 + \ell^2 m^2},
\ee
shifts the singular points, transforming the radial equation~\eqref{eq:radial} into the canonical hypergeometric equation
\be\label{eq:hypergeomEq}
    z(1 - z)\frac{d^2\phi(z)}{dz^2} + (c - (a + b + 1)z)\frac{d\phi(z)}{dz} - ab\,\phi(z) = 0,
\ee
where
\be\label{eq:hypergeomParam}
    z = \frac{r_h^2}{r^2}, \quad a, b = \frac{1}{2}\left(1 + \nu \mp \frac{i\ell J}{r_h} + \frac{i\ell^2\om}{r_h}\right), \quad c = 1 + \nu.
\ee
The general solution to~\eqref{eq:hypergeomEq} is given in terms of hypergeometric functions~\cite{AbramowitzStegun1965}.

For the case of a massive field, none of the quantities~$c$, ${c - a - b}$ and ${a - b}$ are integers. Therefore, the general solution near the boundary (${r \to \infty}$) has the form
\be
    \begin{aligned}
        \phi(r)_{m\,\neq\,0} &= \left(1 - \frac{r_h^2}{r^2}\right)^{\frac{i\ell^2\om}{2r_h}}\left[\const_1\left(\frac{r_h}{r}\right)^{\frac{1}{2} + \nu}\,\HyperF{a}{b}{c}{\frac{r_h^2}{r^2}}\right. +\\
        & + \left.\const_2 \left(\frac{r_h}{r}\right)^{\frac{1}{2} - \nu}\HyperF{a - c + 1}{b - c + 1}{2 - c}{\frac{r_h^2}{r^2}}\right],
    \end{aligned}
    \label{eq:BTZ_GenSol_massive}
\ee
with integration constants~$\const_1$ and~$\const_2$. Since the hypergeometric function is symmetric in its first two arguments, the solution is symmetric with respect to inversion of the sign of~$J$.

The massless case should be considered separately since the parameter ${c = 2}$ is integer. The first solution has the same form as in the massive case, however, the second solution is
\be
    \begin{aligned}
        f_2(r)_{m\,=\,0} &= \const_2\left(1 - \frac{r_h^2}{r^2}\right)^{\frac{i\ell^2\om}{2r_h}}\sqrt{\frac{r}{r_h}}\left[\ln\left(\frac{r_h^2}{r^2}\right)\HyperF{a}{b}{2}{\frac{r_h^2}{r^2}} + \frac{1}{(a - 1)(b - 1)}\frac{r^2}{r_h^2}\right. + \\
        & + \sum_{s\,=\,1}^{\infty} \frac{(a)_s(b)_s}{s!(2)_s} \left(\frac{r_h}{r}\right)^{2s} \left(\psi(a + s) - \psi(a) + \psi(b + s) - \psi(b)\right. - \\
        & - \left.\left.\psi(2 + s) + \psi(2) - \psi(1 + s) +  \psi(1)\right)\right],
    \end{aligned}
\ee
where $(a)_s$ is the Pochhammer symbol, $\psi(z)$ is the digamma function and
\be
    a, b = 1 \mp \frac{i\ell J}{2r_h} + \frac{i\ell^2\om}{2r_h}.
\ee

The large-distance asymptotic behavior of the second solution is $r^{\nu - 1/2}$ for ${m \neq 0}$ and $\sqrt{r}\ln(r)$ for ${m = 0}$, since the hypergeometric function is constant to the leading order. The boundary condition at ${r \to \infty}$ hence requires to set ${\const_2 = 0}$. The other constant then becomes just an overall multiplicative factor that can be chosen arbitrarily and we fix it as ${\const_1 = 1}$. The solution near the horizon can be obtained by transforming the hypergeometric function between the vicinities of the singular points,
\footnote{
    We use a linear transformation formula~\cite{AbramowitzStegun1965},   
    \be
        \begin{aligned}
            & \HyperFTransformed{a}{b}{c}{z} = \frac{\Gamma(c)\Gamma(c - a - b)}{\Gamma(c - a)\Gamma(c - b)}\HyperF{a}{b}{a + b + 1 - c}{1 - z} + \\
            & + \frac{\Gamma(c)\Gamma(a + b - c)}{\Gamma(a)\Gamma(b)}\left(1 - z\right)^{c - a - b}\HyperF{c - a}{c - b}{1 + c - a - b}{1 - z},
        \end{aligned}
    \ee
    to ensure that the solution that involves hypergeometric function is well-behaved for the argument close to~$1$ which is important since the radius of convergence of the hypergeometric function $\HyperF{a}{b}{c}{z}$ is unity, ${|z| < 1}$.
}
\be
    \begin{aligned}
        \phi(r) = 
        & \left(1 - \frac{r_h^2}{r^2}\right)^{\frac{i\ell^2\om}{2r_h}}\left(\frac{r_h}{r}\right)^{\frac{1}{2} + \nu}\left[\frac{\Gamma(c)\Gamma(c - a - b)}{\Gamma(c - a)\Gamma(c - b)}\HyperF{a}{b}{a + b + 1 - c}{1 - \frac{r_h^2}{r^2}} + \right. \\
        & + \left.\frac{\Gamma(c)\Gamma(a + b - c)}{\Gamma(a)\Gamma(b)}\left(1 - \frac{r_h^2}{r^2}\right)^{c - a - b}\HyperF{c - a}{c - b}{1 + c - a - b}{1 - \frac{r_h^2}{r^2}}\right].
    \end{aligned}
\ee
Since $\overline{c - a} = a$, $\overline{c - b} = b$, $\overline{a + b + 1 - c} = 1 + c - a - b$, $\overline{a + b - c} = c - a - b$ and $c - a - b = -i\ell^2\om/r_h$, this function is real-valued.

The remaining boundary condition  then gives an equation ${\phi(r_0) = 0}$ which we solve numerically with respect to~$\om$. To set up the values of the stretched horizon location close enough to the actual horizon, we  use the tortoise coordinate~$z$,
\be
    z = \frac{\ell^2}{r_h}\arctanh\left(\frac{r_h}{r}\right), \quad r > r_h.
    \label{eq:BTZ_tort}
\ee
In the near-horizon limit, ${r_0 = r_h + \eps}$ with $\eps \to 0$, the stretched horizon corresponds to a sufficiently large value~$z_0$ of the tortoise coordinate. In this case, the boundary condition takes a particularly simple form for $r_h = 1$ and $\ell = 1$,
\be\label{eq:horEq}
    \begin{aligned}
        & \frac{\eps^{\frac{i\om}{2}}}{\Gamma(1 + i\om)\Gamma\left(\frac{1}{2}(1 + \nu + i(J - \om))\right)\Gamma\left(\frac{1}{2}(1 + \nu - i(J + \om))\right)} = \\ 
        & = \frac{\eps^{-\frac{i\om}{2}}}{\Gamma(1 - i\om)\Gamma\left(\frac{1}{2}(1 + \nu - i(J - \om))\right)\Gamma\left(\frac{1}{2}(1 + \nu + i(J + \om))\right)}.
    \end{aligned}
\ee
In the following, we will keep ${\ell = 1}$ for simplicity.

Besides the condition at infinity~\eqref{eq:outer_bc}, the normal modes also arise if the outer Dirichlet boundary condition is imposed at finite radial distance ${r_1 > r_0}$. In this situation, one needs to relate the two integration constants of the general solution using one of the boundary conditions and then solve the remaining one numerically. In the massive case~\eqref{eq:BTZ_GenSol_massive}, this relation is
\be
    \const_1 = -\const_2\left(\frac{r_1}{r_h}\right)^{2\nu}\frac{\HyperF{a - c + 1}{b - c + 1}{2 - c}{\frac{r_h^2}{r_1^2}}}{\HyperF{a}{b}{c}{\frac{r_h^2}{r_1^2}}},
\ee
Choosing the second constant as ${\const_2 = -(r_h/r_1)^{2\nu}\HyperF{a}{b}{c}{r_h^2/r_1^2}}$ we obtain the solution near the outer boundary, ${r, r_1 \gg r_h}$,
\be
    \begin{aligned}
        \phi(r)_{m\,\neq\,0} &= \left(1 - \frac{r_h^2}{r^2}\right)^{\frac{i\ell^2\om}{2r_h}}\left[\left(\frac{r_h}{r}\right)^{\frac{1}{2} + \nu}\HyperF{a - c + 1}{b - c + 1}{2 - c}{\frac{r_h^2}{r_1^2}}\HyperF{a}{b}{c}{\frac{r_h^2}{r^2}}\right. -\\
        & - \left. \left(\frac{r_h}{r}\right)^{\frac{1}{2} - \nu}\left(\frac{r_h}{r_1}\right)^{2\nu}\,\HyperF{a}{b}{c}{\frac{r_h^2}{r_1^2}}\HyperF{a - c + 1}{b - c + 1}{2 - c}{\frac{r_h^2}{r^2}}\right].
    \end{aligned}
\ee
To get the solution near the horizon, those hypergeometric functions that depend on~$r$ should be transformed yielding the two following parts of the solution,
\be
    \begin{aligned}
        f_1(r) &= \left(1 - \frac{r_h^2}{r^2}\right)^{\frac{i\ell^2\om}{2r_h}}\left(\frac{r_h}{r}\right)^{\frac{1}{2} + \nu}\HyperF{a - c + 1}{b - c + 1}{2 - c}{\frac{r_h^2}{r_1^2}}\HyperFTransformed{a}{b}{c}{\frac{r_h^2}{r^2}}, \\
        f_2(r) &= \left(1 - \frac{r_h^2}{r^2}\right)^{\frac{i\ell^2\om}{2r_h}}\left(\frac{r_h}{r}\right)^{\frac{1}{2} - \nu}\left(\frac{r_h}{r_1}\right)^{2\nu}\,\HyperF{a}{b}{c}{\frac{r_h^2}{r_1^2}}\HyperFTransformed{a - c + 1}{b - c + 1}{2 - c}{\frac{r_h^2}{r^2}}, \\
        & \HyperFTransformed{a}{b}{c}{z} = \frac{\Gamma(c)\Gamma(c - a - b)}{\Gamma(c - a)\Gamma(c - b)}\HyperF{a}{b}{a + b + 1 - c}{1 - z} + \\
        & + \frac{\Gamma(c)\Gamma(a + b - c)}{\Gamma(a)\Gamma(b)}\left(1 - z\right)^{c - a - b}\HyperF{c - a}{c - b}{1 + c - a - b}{1 - z}.
    \end{aligned}
\ee

Numerical analysis shows that the shifting of $r_1$-boundary has no effect on the spectrum since the zeros of~$f_1$ and~$f_2$ with respect to~$\om$ coincide. This means that the shape of the SFF is determined only by the position of the inner boundary condition and not by the outer one.

%----------------------%
\subsection{SFF details}

By analogy with quantum mechanics, normal modes $\om_{n,J}$ can be interpreted as energy eigenvalues of some quantum mechanical system~\cite{tHooft:1984kcu, Ichinose:1994rg, Das:2022evy} at finite temperature~$1/\beta$. This gives an analytically continued partition function
\be
    Z(\beta, t) = \sum\limits_{n\,=\,1}^{n\cut}\sum\limits_{J\,=\,-J\cut}^{J\cut} e^{-(\beta - it)\,\om_{n,J}},
    \label{eq:Z}
\ee
where the summation is performed up to finite cutoff values. Then, the spectral form factor (SFF) is defined as~\cite{Cotler:2016fpe}
\be
    g(\beta, t) = \frac{\abs{Z(\beta, t)}^2}{\abs{Z(\beta, 0)}^2}.
    \label{eq:SFF}
\ee
This quantity takes into account the correlations between any two energy levels of the system. Indeed, its infinite temperature version can be obtained as the Fourier transform of the two-level correlation function $\rho^{(2)}(\lambda_i, \lambda_j)$~\cite{Liu:2018hlr}, 
\be
    g(t) = \sum_{i,\,j\,=\,1}^{N}\int d\lambda_i d\lambda_j\,\rho^{(2)}(\lambda_i, \lambda_j) e^{-i(\lambda_i - \lambda_j)t},
    \label{eq:SFF_FT}
\ee
for a spectrum consisting of $N$ levels. The correlation function here is a reduced joint probability density function, defined as
\be
    \rho^{(2)}(\lambda_i, \lambda_j) = \int d\lambda_1 \dots d\lambda_{i - 1}d\lambda_{i + 1} \dots d\lambda_{j - 1}d\lambda_{j + 1} \dots d\lambda_N P(\lambda_1, \dots, \lambda_N),
    \label{eq:corrfunc}
\ee
where $P(\lambda_1, \dots, \lambda_N)$ is the joint probability density function. The particular form of~$P$ is a subject to the ensemble distribution; however, if there is only a single realization of the system, Fourier transform~\eqref{eq:SFF_FT} reduces trivially to the summation over levels as in~(\ref{eq:Z},~\ref{eq:SFF}).

Note that the system is supposed to be thermal but the temperature~$1/\beta$ is not related to the Hawking temperature associated with the horizon. Indeed, the scalar field that we study here does not model Hawking radiation but is rather an independent field on the curved background.

It was shown in~\cite{Das:2022evy} that a dip-ramp-plateau structure of the SFF for the normal modes of massless scalar field occurs when~$r_0$ is slightly larger than~$r_h$. In previous papers~\cite{Das:2022evy, Das:2023ulz, Krishnan:2023jqn}, the discussions were focused on the massless case, where the DRP structure of the form factor arose with even sufficiently small values of~$J\cut$ and~$n\cut$. 

\begin{figure}[t]\centering
    \subfloat[massless]{\includegraphics[width=0.33\linewidth]{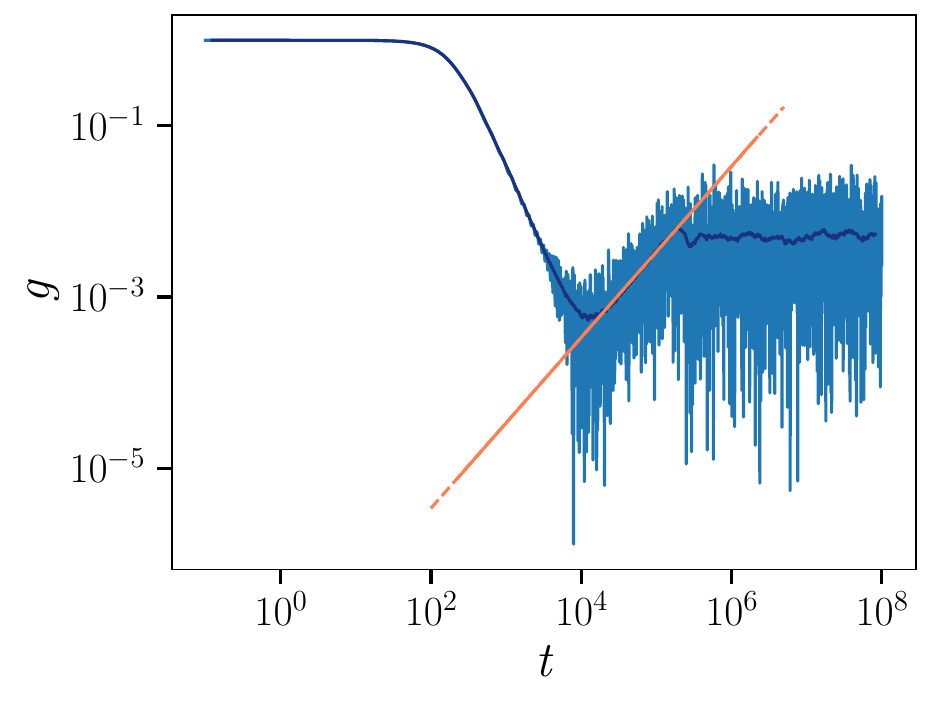}}
    \subfloat[$m = 30$]{\includegraphics[width=0.33\linewidth]{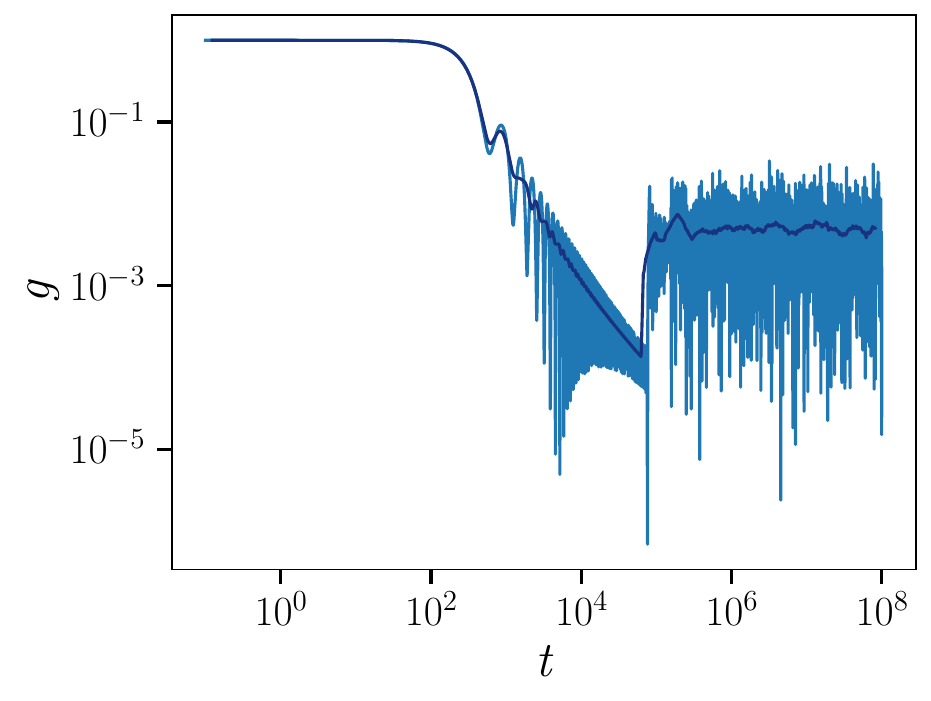}}
    \subfloat[$m = 60$]{\includegraphics[width=0.33\linewidth]{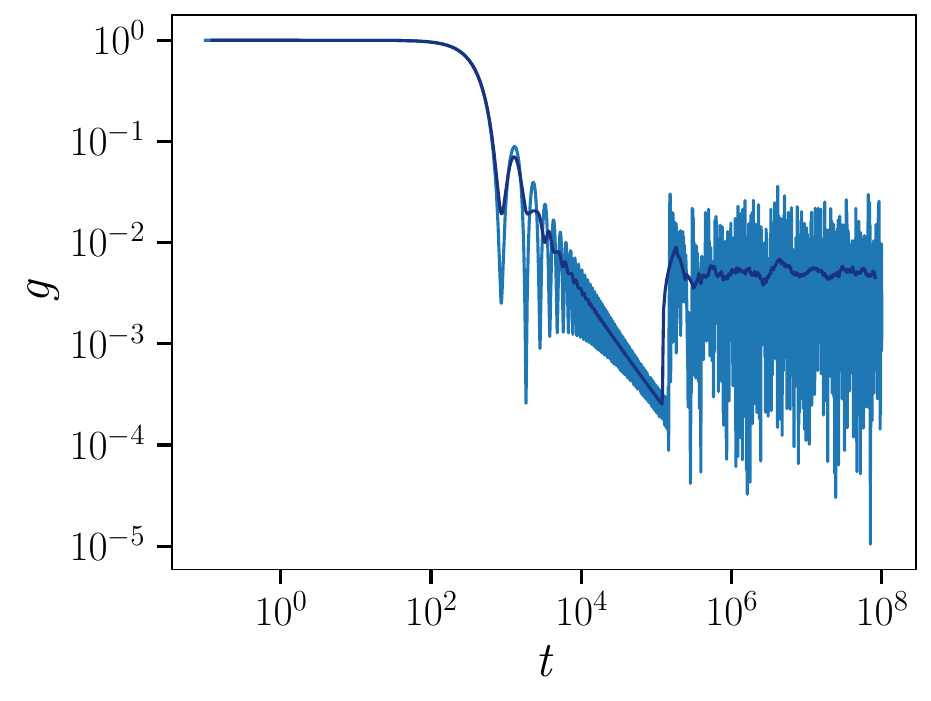}}
    \caption{Mass dependence of the SFF for the BTZ black hole case with ${J\cut = 200}$, ${n\cut = 1}$, ${r_h = 1}$, ${z_0 = 30}$ and ${\beta = 0}$. With a darker line, the moving averaged data is shown. Dashed orange line marks a linear fit of the averaged data for the ramp part while solid orange line with a unit slope is plotted for reference.}
    \label{fig:BTZ_mass_break}
\end{figure}

\begin{figure}[t]\centering
    \subfloat[massless]{\includegraphics[width=0.33\linewidth]{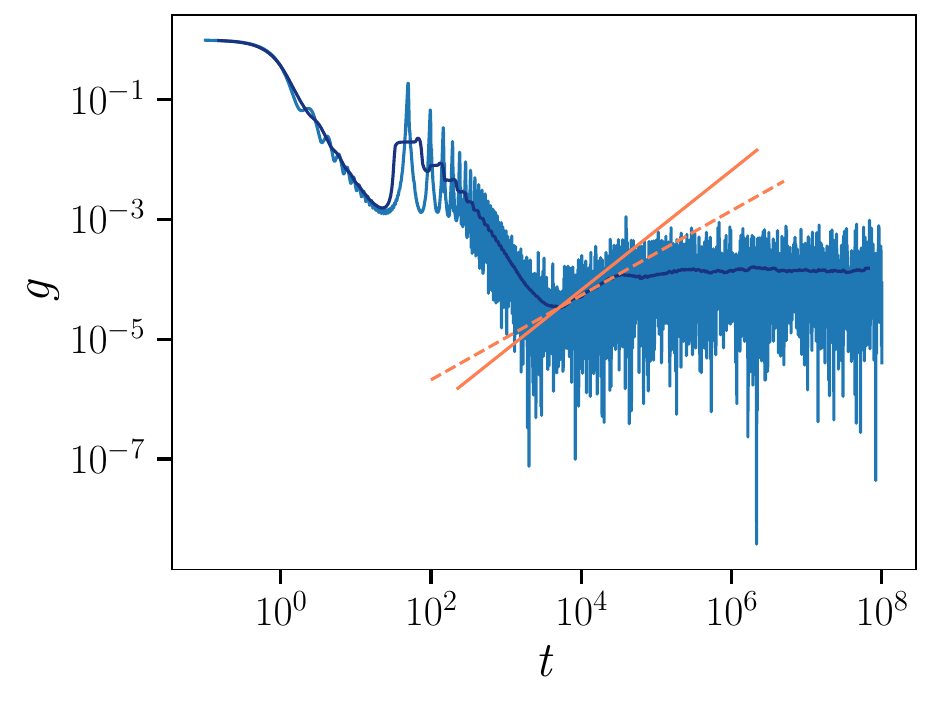}}
    \subfloat[$m = 30$]{\includegraphics[width=0.33\linewidth]{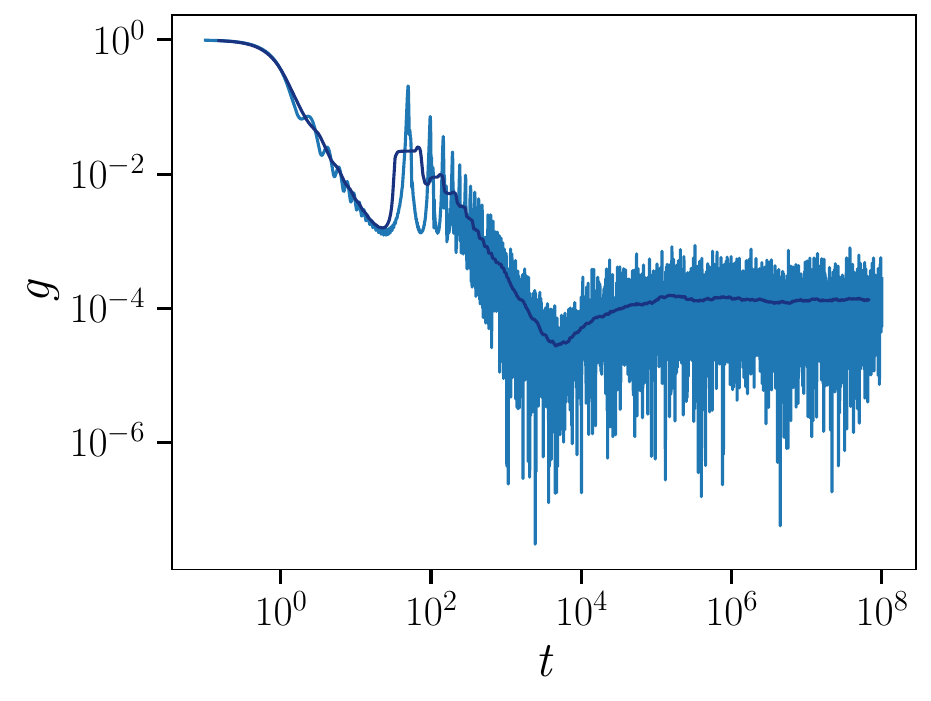}}
    \subfloat[$m = 60$]{\includegraphics[width=0.33\linewidth]{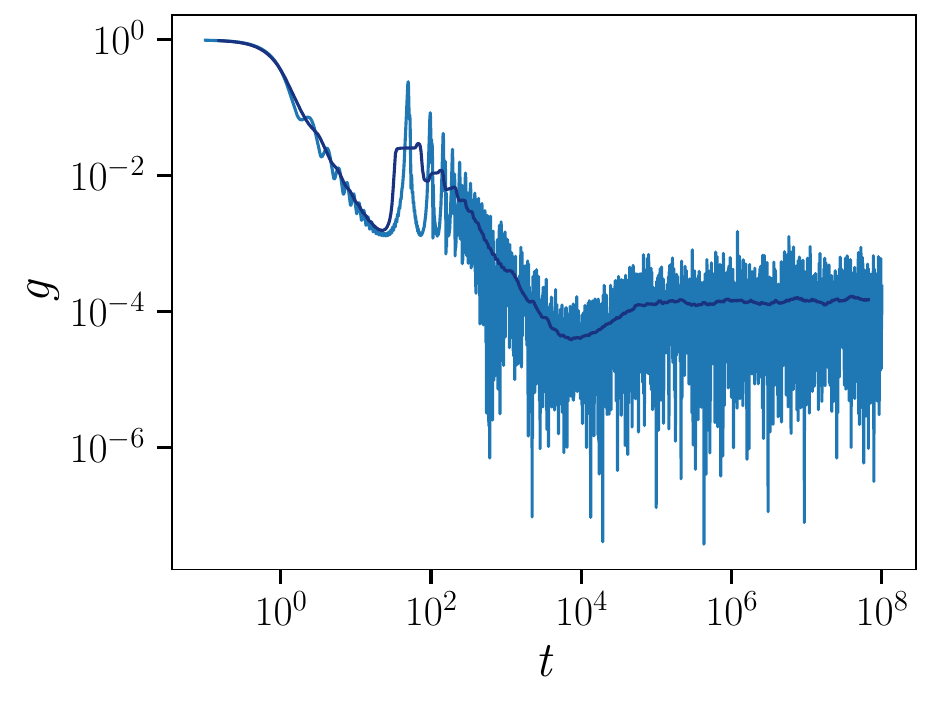}}
    \caption{Recovering the DRP structure of the SFF for the massive field at the BTZ black hole background with ${J\cut = 300}$, ${n\cut = 30}$, ${r_h = 1}$, ${z_0 = 30}$ and ${\beta = 0.5}$. With a darker line, the moving averaged data is shown. Dashed orange line marks a linear fit of the averaged data for the ramp part while solid orange line with a unit slope is plotted for reference.}
    \label{fig:BTZ_mass}
\end{figure}

However, in the massive case, it is necessary to consider larger values of~$J\cut$ and~$n\cut$ in order to reproduce the DRP structure. \figref{fig:BTZ_mass_break} shows how the SFF changes with increasing mass of the scalar field. The graphs are plotted for the other parameters being fixed. As the mass increases, the transition from the dip to the plateau in the dip-ramp-plateau structure becomes more abrupt, and with a large enough mass, the ramp is almost absent. We also found that as the stretched horizon gets closer to the actual horizon, a clear dip-ramp-plateau structure remains at larger mass. If the stretched horizon~$z_0$ is fixed, the DRP structure also improves as~$J\cut$ and~$n\cut$ increase (see \figref{fig:BTZ_mass}).

The discreteness of the energy spectrum along with the absence of ensemble averaging cause erratic oscillations starting from intermediate times but it can be moving averaged to reveal the large-scale details of the SFF. The dashed line in Figures~\ref{fig:BTZ_mass_break} and~\ref{fig:BTZ_mass} represents a linear fit to moving average of the ramp part. Based on this, we can see that the ramp part has a slope close to unit, which is indicated by the solid orange line.

It was noted in~\cite{Das:2022evy}, that moving the inner boundary condition out of the actual horizon leads to the gradual vanishing of the dip-ramp-plateau structure. It is also possible to consider the structure of the form factor when the size of the horizon is enlarged. If one fixes the tortoise coordinate~$z_0$, the difference between the radial coordinate~$r_0$ and~$r_h$ will decrease with the growth of~$r_h$ according to~\eqref{eq:BTZ_tort}. In this case, the DRP structure transforms in the same way as when mass is changed: as~$r_h$ increases, the slope of the ramp becomes steeper and leads to a sudden transition between the dip and the plateau.

%-----------------------%
\section{\label{sec:dS}De Sitter}
\subsection{Scalar field on de Sitter background}

In this section, we consider a scalar field $\Phi$ of mass $m$ on the ${2 + 1}$-dimensional de Sitter background,which metric in the static coordinates has the form
\be
   ds^2_{\text{dS}} = -\left(1 - \frac{r^2}{r_c^2}\right)dt^2 + \frac{dr^2}{1 - \frac{r^2}{r_c^2}} + r^2d\varphi^2,
   \label{eq:dS_static}
\ee
where the cosmological horizon is located at ${r = r_c}$ and we consider the region inside the horizon ${0 < r < r_c}$.

Again, we are interested in solutions of the massive Klein-Gordon equation. The explicit solution for minimally-coupled massive scalar field in the static patch of ${2 + 1}$-dimensional de Sitter was obtained in~\cite{Abdalla:2002rm} in the context of quasi-normal mode analysis. Following the previous section, in order to obtain the normal modes we impose a Dirichlet boundary condition ${\phi(r_0) = 0}$ for the stretched horizon located at finite ${r_0 < r_c}$ and also require that the solution is not divergent at ${r = 0}$.

With the same form of $\Phi$~\eqref{eq:ansatz} as for the BTZ case, the radial part of the equation satisfies
\be\label{eq:RadEoM}
    (r^2 - r_c^2)^2\frac{d^2\phi(r)}{dr^2} + 2r(r^2 - r_c^2)\frac{d\phi(r)}{dr} + r_c^4\,\om^2\phi(r) - V(r)\phi(r) = 0,
\ee
where 
\be
    V(r) = (r_c^2 - r^2)\Bigg[\frac{r_c^2}{r^2}\left(J^2 - \frac{1}{4}\right) + r_c^2\,m^2 - \frac{3}{4}\Bigg].
\ee
In this case, the ansatz transforming the radial equation with three singular points into the canonical hypergeometric equation~\eqref{eq:hypergeomEq} has a form
\be
    \phi(r) \to \left(1 - \frac{r^2}{r_c^2}\right)^\alpha \left(\frac{r}{r_c}\right)^\beta \phi\left(\frac{r^2}{r_c^2}\right),
\ee
with
\be
    \alpha = \frac{i r_c\,\om}{2}, \quad \beta = \frac{1}{2} + \abs{J}.
\ee

In the resulting equation, the independent variable is ${z = r^2/r_c^2}$ and the coefficients are
\be
    a, b = \frac{1}{2} \left(1 \mp\sqrt{1 - r_c^2 m^2} + i r_c\,\om + \abs{J}\right), \quad c = 1 + \abs{J}.
\ee
Note, that the parameter~$c$ takes integer values since $J$~are integers, which leads to a special form of the second solution~\cite{AbramowitzStegun1965}. The first solution is
\be\label{eq:dS_Phi}
    f_1(r) = \const_1\left(1 - \frac{r^2}{r_c^2}\right)^{\frac{i r_c \om}{2}}\left(\frac{r}{r_c}\right)^{\frac{1}{2} +\,\abs{J}}\,\HyperF{a}{b}{1 + \abs{J}}{\frac{r^2}{r_c^2}},
\ee
while the second solution for ${J \neq 0}$ is
\be
    \begin{aligned}
        f_2(r)_{J\,\neq\,0} &= \const_2\left(1 - \frac{r^2}{r_c^2}\right)^{\frac{i r_c \om}{2}}\left(\frac{r}{r_c}\right)^{\frac{1}{2} + \,\abs{J}}\left[\ln\left(\frac{r^2}{r_c^2}\right)\HyperF{a}{b}{1 + \abs{J}}{\frac{r^2}{r_c^2}}\right. - \\
        & - \sum\limits_{s\,=\,1}^{\abs{J}} \frac{(s - 1)!(-\abs{J})_s}{(1 - a)_s(1 - b)_s}\left(\frac{r}{r_c}\right)^{-2s} + \\ 
        & + \sum\limits_{s\,=\,1}^\infty\frac{(a)_s(b)_s}{(1 + \abs{J})_s s!}\left(\frac{r}{r_c}\right)^{2s}(\psi(a + s) - \psi(a) + \psi(b + s) - \psi(b) - \\
        & - \left.\psi(1 + \abs{J} + s) + \psi(1 + \abs{J}) - \psi(1 + s) + \psi(1))\right],
    \end{aligned}
\ee
and for ${J = 0}$ is
\be
    \begin{aligned}
        & f_2(r)_{J\,=\,0} = \const_2\left(1 - \frac{r^2}{r_c^2}\right)^{\frac{i r_c \om}{2}}\sqrt{\frac{r}{r_c}}\left[\ln\left(\frac{r^2}{r_c^2}\right)\HyperF{a}{b}{1}{\frac{r^2}{r_c^2}}\right. + \\ 
        & + \left.\sum\limits_{s\,=\,1}^\infty\frac{(a)_s(b)_s}{(s!)^2}\left(\frac{r}{r_c}\right)^{2s}(\psi(a + s) - \psi(a) + \psi(b + s) - \psi(b) - 2\psi(1 + s) + 2\psi(1))\right],
    \end{aligned}
\ee
where $(a)_s$ is the Pochhammer symbol, $\psi(z)$ is the digamma function and $\const_{1,2}$ are integration constants.

Since in both cases the function~$f_2$ is divergent at ${r \to 0}$ due to the logarithmic contribution, a normalization condition at the origin implies ${\const_2 = 0}$. We also fix the remaining constant as ${\const_1 = 1}$. Transforming the hypergeometric function in~\eqref{eq:dS_Phi}, we obtain a solution near the horizon,
\be
    \begin{aligned}
        \phi(r) = 
        & \left(1 - \frac{r^2}{r_c^2}\right)^{\frac{i r_c \om}{2}}\left(\frac{r}{r_c}\right)^{\frac{1}{2} + \,\abs{J}}\left[\frac{\Gamma(c)\Gamma(c - a - b)}{\Gamma(c - a)\Gamma(c - b)}\,\HyperF{a}{b}{a + b + 1 - c}{1 - \frac{r^2}{r_c^2}} + \right. \\
        & + \left.\frac{\Gamma(c)\Gamma(a + b - c)}{\Gamma(a)\Gamma(b)}\left(1 - \frac{r^2}{r_c^2}\right)^{c - a - b}\HyperF{c - a}{c - b}{1 + c - a - b}{1 - \frac{r^2}{r_c^2}}\right],
    \end{aligned}
\ee
which is a real-valued function.

The tortoise coordinate for the de Sitter static patch~\eqref{eq:dS_static} is
\be
    z = r_c \arctanh\left(\frac{r}{r_c}\right),
\ee
for which~$r_0$ slightly less than~$r_c$ corresponds to a large enough~$z_0$. Applying the boundary condition at the stretched horizon, ${\phi(z_0) = 0}$, we find a discrete set of normal modes~$\om_{n,J}$. In the near-horizon approximation $r = r_c - \eps$ with ${\eps \to 0}$ and ${r_c = 1}$, this equation simplifies to 
\be
    \begin{aligned}
        & \frac{\eps^{\frac{i\om}{2}}\Gamma(-i\om)}{\Gamma\left(\frac{1}{2}(1 - \sqrt{1 - m^2} + J - i\om))\right)\Gamma\left(\frac{1}{2}(1 + \sqrt{1 - m^2} + J - i\om))\right)} = \\ 
        & = \frac{\eps^{-\frac{i\om}{2}}\Gamma(i\om)}{\Gamma\left(\frac{1}{2}(1 - \sqrt{1 - m^2} + J + i\om))\right)\Gamma\left(\frac{1}{2}(1 + \sqrt{1 - m^2} + J + i\om))\right)}.
    \end{aligned}
\ee

%-----------------------%
\subsection{SFF for de Sitter background}

\begin{figure}[t]\centering
    \subfloat[massless, $n\cut = 1$, $\beta = 0$]{\includegraphics[width=0.33\linewidth]{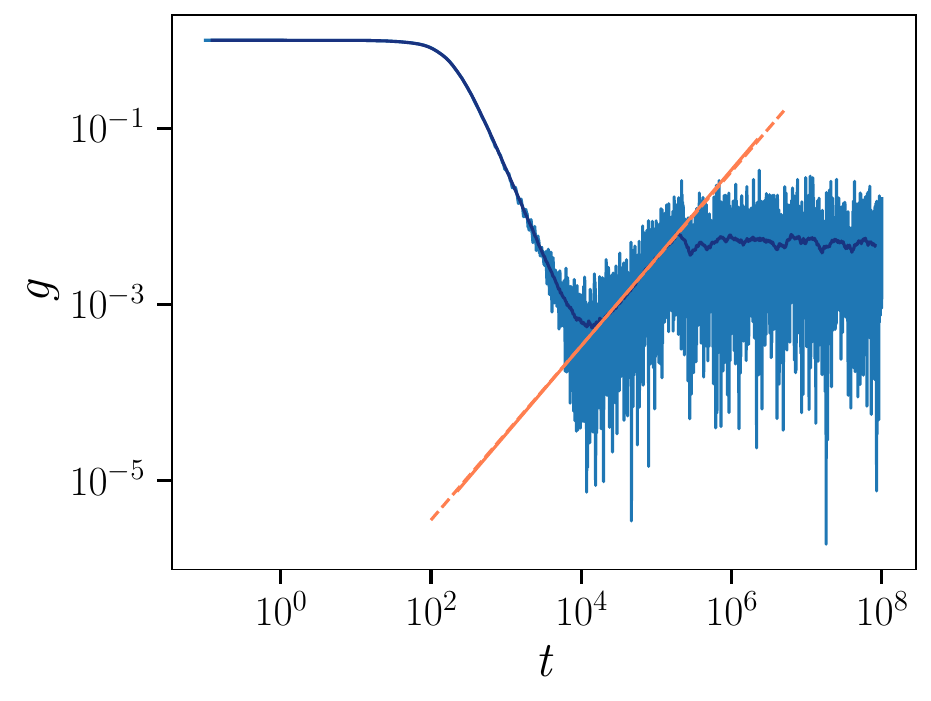}}
    \subfloat[$m = 30, n\cut = 1$, $\beta = 0$]{\includegraphics[width=0.33\textwidth]{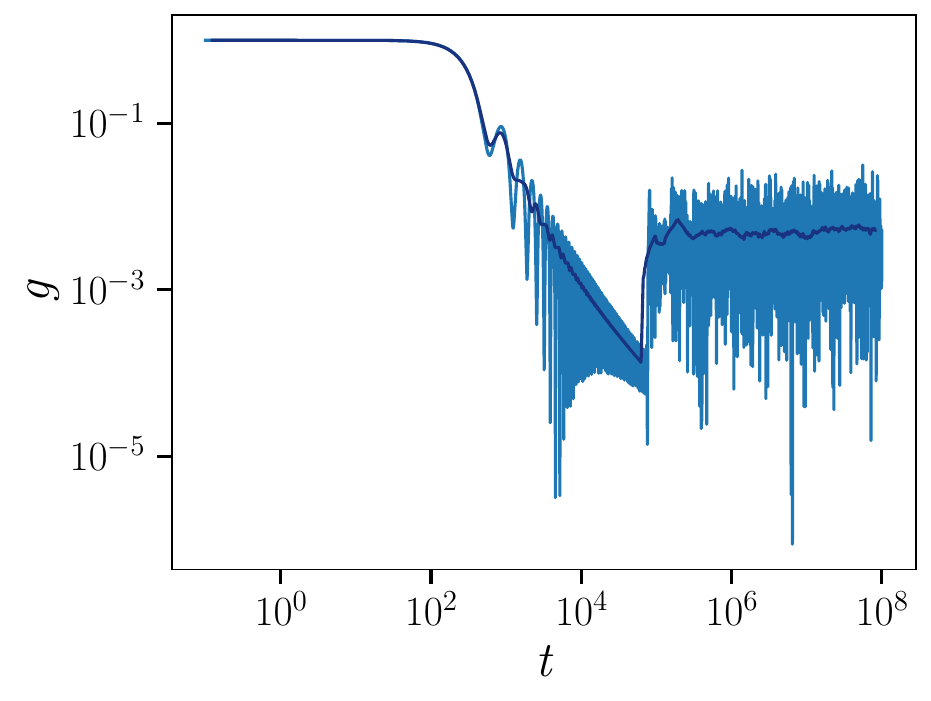}}
    \subfloat[$m = 30, n\cut = 30$, $\beta = 10$]{\includegraphics[width=0.33\linewidth]{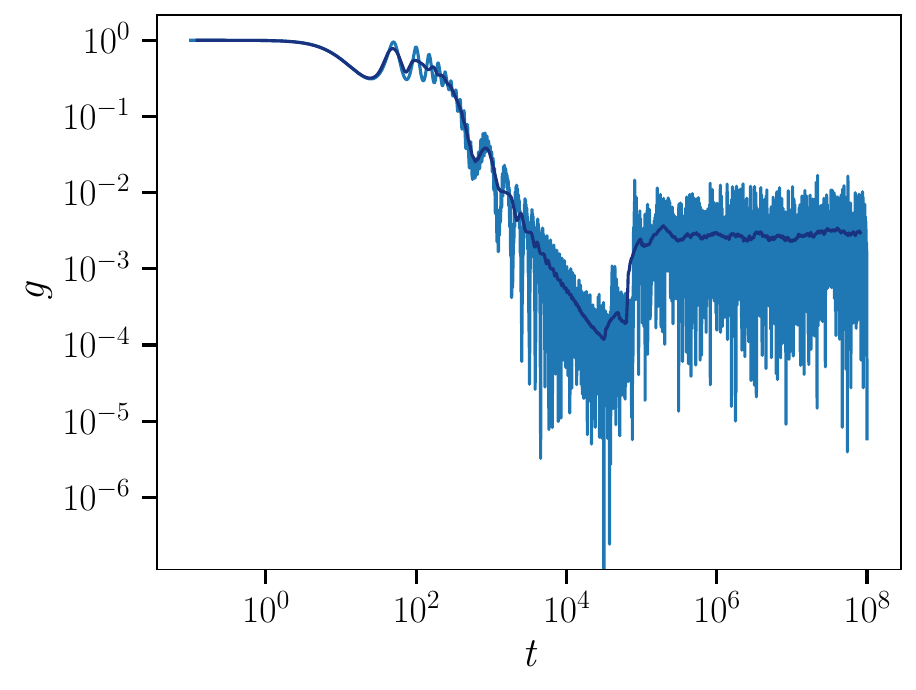}}
    \caption{Mass dependence of the SFF for the de Sitter case with ${J\cut = 200}$, ${r_c = 1}$, ${z_0 = 30}$. With a darker line, the moving averaged data is shown. Dashed orange line marks a linear fit of the averaged data for the ramp part while solid orange line with a unit slope is plotted for reference.}
    \label{fig:dS_mass}
\end{figure}

Once the spectrum is known, calculation of the spectral form factor can be done straightforwardly, drawing an analogy with the BTZ black hole. In \figref{fig:dS_mass}, we show the SFF for the de Sitter background. Similarly to the case of the BTZ black hole, this graph also indicates the dip-ramp-plateau structure. The slope of the ramp becomes stable as~$J\cut$ increases, so it would be sufficient to choose ${J\cut = 200}$ in numerical calculations. Increasing the mass ruins the dip-ramp-plateau structure of SFF, the ramp on the middle panel (${m = 30}$) is almost absent. However, at a fixed stretched horizon, increasing the cutoffs recovers the structure as shown in the right panel.

\begin{figure}[t]\centering
    \subfloat[$\beta = 0$]{\includegraphics[width=0.33\linewidth]{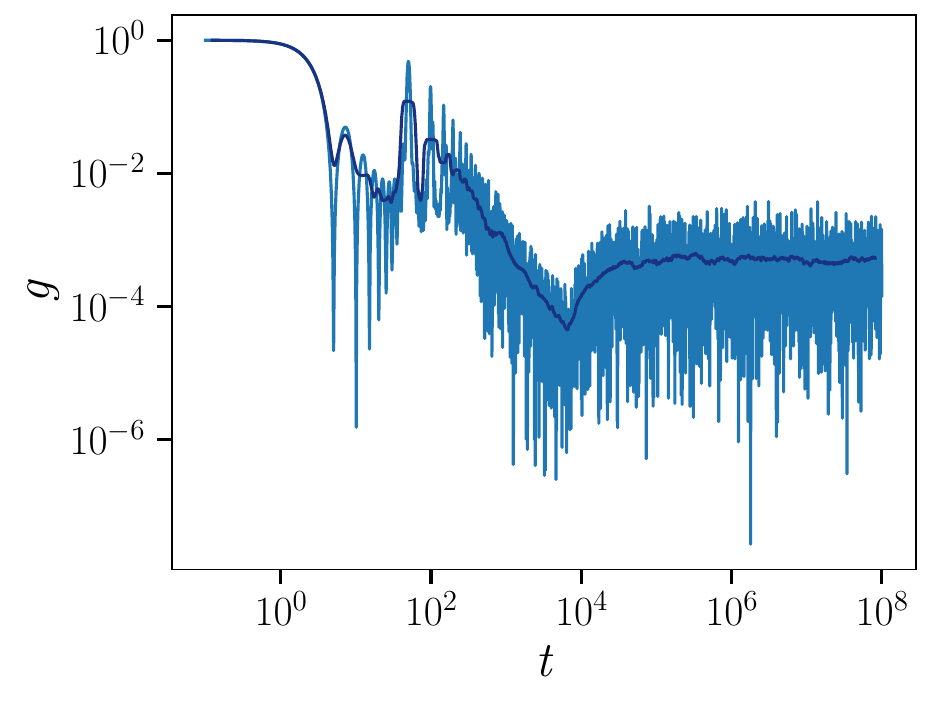}}
    \subfloat[$\beta = 1$]{\includegraphics[width=0.33\textwidth]{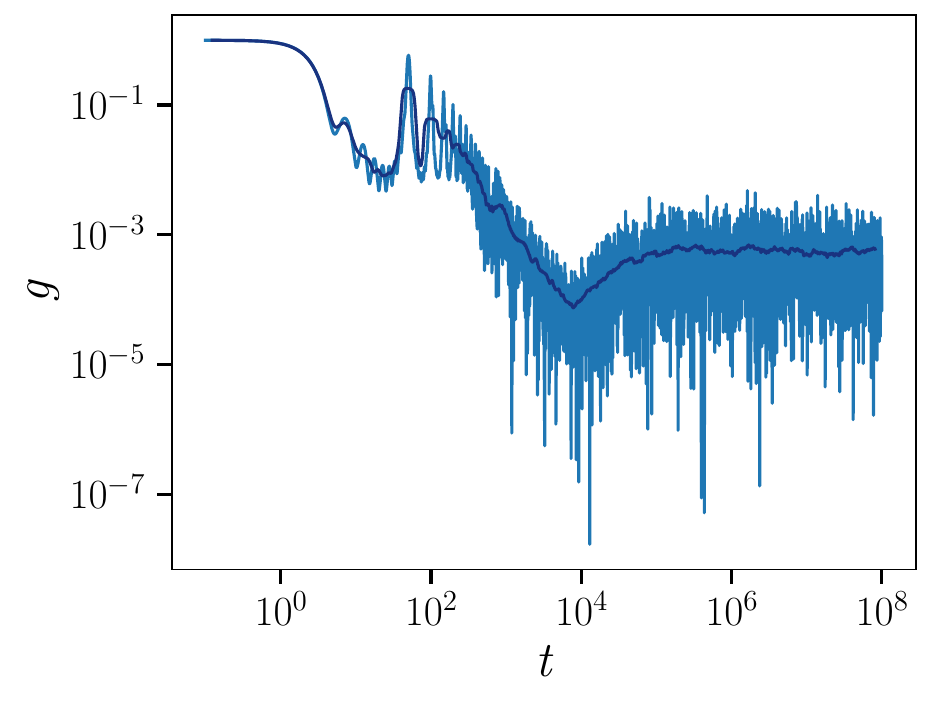}}
    \subfloat[$\beta = 30$]{\includegraphics[width=0.33\textwidth]{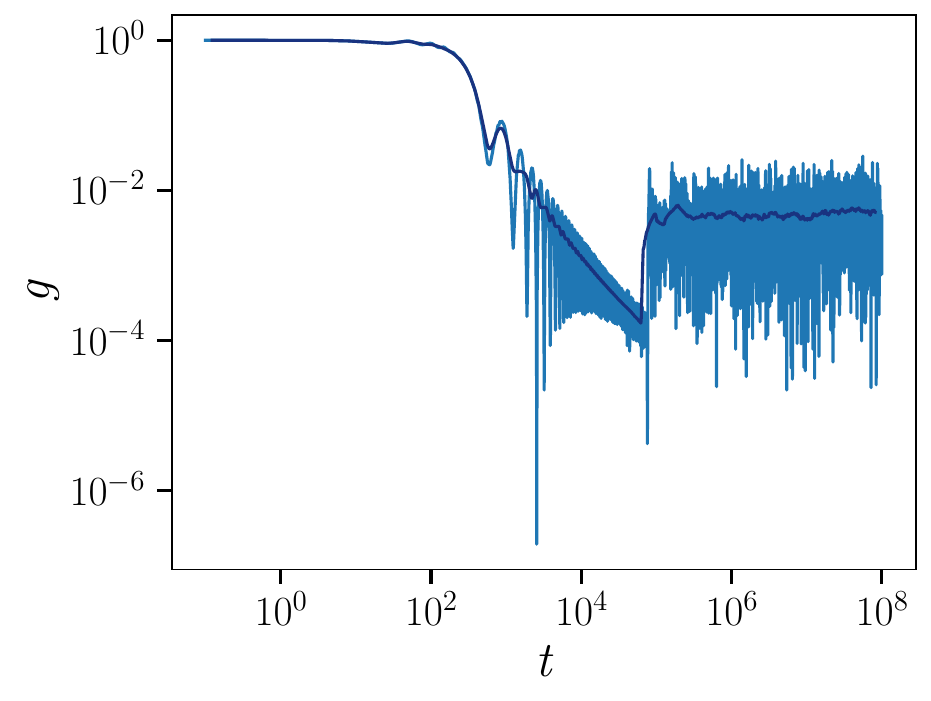}}
    \caption{$\beta$-dependence of the SFF for the de Sitter case with $m = 30$, $J\cut = 200$, $n\cut = 10$, $r_c = 1$, $z_0 = 30$. With a darker line, the moving averaged data is shown.}
    \label{fig:dS_beta}
\end{figure}

The dependence of the SFF on the inverse temperature is shown in \figref{fig:dS_beta}. As in the case of the BTZ background~\cite{Das:2022evy}, the fluctuations of the dip part of the SFF occur for small values of~$\beta$ and ${n\cut > 1}$. As~$\beta$ increases, the dip becomes smoother. At ${n\cut = 1}$, these fluctuations are absent and the dip-ramp-plateau structure is observed regardless of the value of~$\beta$.

\begin{figure}[t]\centering
    \includegraphics[width=0.4\linewidth]{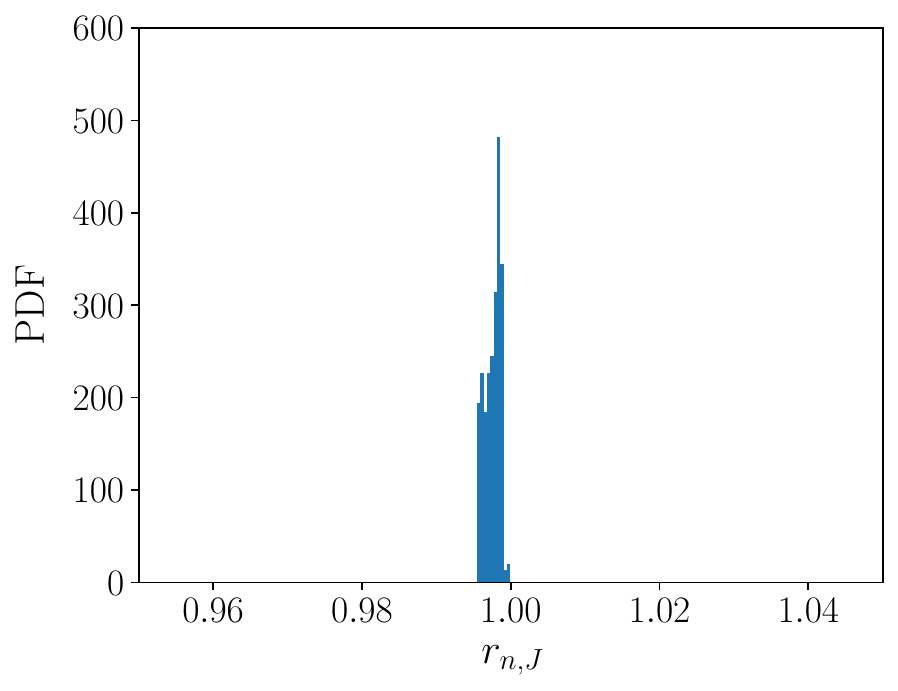}
    \caption{The probability density function (PDF) of the ratios of level spacings for the de Sitter case with ${r_c = 1}$, ${z_0 = 30}$, ${\beta = 0}$ and ${m = 0}$. The number of modes taken into account is chosen as ${J\cut = 200}$ and ${n\cut = 30}$.}
    \label{fig:dS_LSD}
\end{figure}

Another spectral characteristic used as a signature of chaos in quantum systems is the level spacing distribution~\cite{Bohigas84, Haake:2010fgh}. To avoid the necessity of unfolding procedure, one can consider ratios of spacings instead of spacings themselves~\cite{Oganesyan:2007wpd, Atas:2013gvn}. By level spacings we mean here the differences between the closest relative to~$n$ modes for every fixed value of~$J$, hence their ratios are $r_{n,J} = (\om_{n + 1,J} - \om_{n,J})/(\om_{n,J} - \om_{n - 1,J})$. Unlike the spectral form factor, this measure does not emphasize chaotic behavior for de Sitter normal modes, as illustrated in \figref{fig:dS_LSD}, since the distributions have narrow-peak shape clearly not coinciding with corresponding distributions for Gaussian ensembles of random matrices. But they are not Poisson either which is typical for integrable models. A similar observation was made for the BTZ case in~\cite{Das:2022evy}.

%--------------------------%
\section{\label{sec:GSFF}Generalized spectral form factor}

To get a better understanding of the energy level distribution, let us consider the generalization of the spectral form factor proposed in~\cite{Wei:2024ujf} called the generalized spectral form factor (gSFF). 

As opposed to the common definition~\eqref{eq:SFF_FT} of the spectral form factor for an ensemble with the joint probability density function $P(\lambda_1, \dots, \lambda_N)$,
\be
    \begin{aligned}
        g(t) &= \int d\lambda \sum \limits_{i,\,j\,=\,1}^N \int d\lambda_1 \dots d\lambda_N P(\lambda_1, \dots, \lambda_N) \delta(\lambda - (\lambda_i - \lambda_j)) e^{-i\lambda t} = \\
        &= \int d\lambda \sum \limits_{i,\,j\,=\,1}^N\int d\lambda_i d\lambda_j\,\rho^{(2)}(\lambda_i, \lambda_j)\delta(\lambda - (\lambda_i - \lambda_j)) e^{-i\lambda t},
    \end{aligned}
    \ee
the generalized spectral form factor~$g^{(2)}(t)$ is possible to define as the Fourier transform of a modified correlation function~$c^{(2)}(\lambda)$ whose definition is based on a standard deviation of the levels~${\sigma(\lambda_i, \lambda_j)}$,
\be
    \begin{aligned}
        g^{(2)}(t) &= \int d\lambda \sum \limits_{i,\,j\,=\,1}^N \int d\lambda_1 \dots d\lambda_N P(\lambda_1, \dots, \lambda_N)\delta(\lambda - \sigma(\lambda_i, \lambda_j))e^{-i\lambda t} = \\
        &= \int d\lambda\,c^{(2)}(\lambda) e^{-i\lambda t}.
    \end{aligned}
\ee

Since only the standard deviation depends on the number of levels in the correlation function, this gives a universal definition for a higher-level spectral form factor. \cite{Wei:2024ujf} provides answers for the GUE for two levels in order to compare them with the SFF, as well as for three levels. In principle, the ordinary spectral form factor can also be extended but this is not of much use for ensembles of random matrices in the case of an odd number of levels~\cite{PhysRevE.55.3886}. 

For a random matrix ensemble, given single realization eigenvalues~$\lambda_i$, the two- and three-level generalized spectral form factor can be expressed in the following way,
\be
    \begin{aligned}
        & g^{(2)}(t) = \frac{1}{N^2}\left\langle\sum\limits_{i,\,j\,=\,1}^N \exp\left(-i\sigma(\lambda_i, \lambda_j)t\right)\right\rangle = \frac{1}{N^2}\left\langle\sum\limits_{i,\,j\,=\,1}^N \exp\left(-i\frac{\abs{\lambda_i - \lambda_j}}{2}t\right)\right\rangle, \\
        & g^{(3)}(t) = \frac{1}{N^3}\left\langle \sum\limits_{i,\,j,\,k\,=\,1}^N \exp(-i\sigma(\lambda_i, \lambda_j, \lambda_k)t)\right\rangle,
    \end{aligned}
\ee
where~$N$ is the matrix dimension and ${\langle . \rangle}$ denotes the ensemble average; ${\sigma(\lambda_i, \lambda_j, \lambda_k)}$ represents the standard deviation of three energy levels,
\be
    \sigma(\lambda_i, \lambda_j, \lambda_k) = \sqrt{\frac{1}{3}\left((\lambda_i - \Lambda)^2 + (\lambda_j - \Lambda)^2 + (\lambda_k - \Lambda)^2\right)}, \quad \Lambda = \frac{\lambda_i + \lambda_j + \lambda_k}{3}.
\ee

In our setup, we consider a single realization of the system hence we do not have the ensemble average. We define the two- and three-level generalized spectral form factor for the known spectrum~$\omega_{n,\,J}$ of normal modes by analogy with the standard spectral form factor~\eqref{eq:SFF} in the case ${\beta = 0}$,
\be
    \begin{aligned}
        & g^{(2)}(t) = g^{(2)}_0\sum_{n,\,J}\,\,\sum_{n',\, J'}\,\exp\left(-i\frac{\abs{\om_{n,J} - \om_{n',J'}}}{2}t\right), \\
        & g^{(3)}(t) = g^{(3)}_0\sum_{n,\,J}\,\,\sum_{n',\, J'}\sum_{n'',\, J''} \exp\big[-i\sigma(\om_{n,J},\, \om_{n',J'},\, \om_{n'',J''})t\big],
    \end{aligned}
\ee
where $g^{(2)}_0$, $g^{(3)}_0$ are normalization constants and we assume that the appropriate cutoffs are chosen for the summations. Introducing a non-zero parameter~$\beta$ is technically possible, however, its meaning for the higher-level gSFF is not as clear as for the ordinary SFF which can be factorized into the absolute value squared of a partition function, that is why we avoid to use it here.

%-------------------%
\subsection{Two-level generalized spectral form factor}
\subsubsection{BTZ black hole and de Sitter}

\begin{figure}[t]\centering
    \subfloat[BTZ]
    {
        \includegraphics[width=\twopanelfigurewidth]{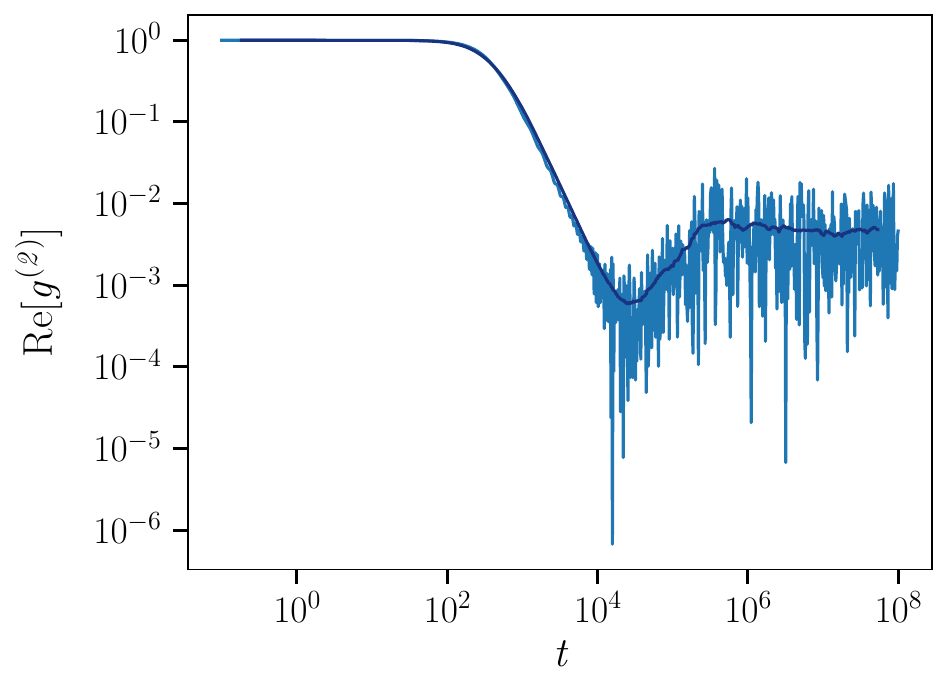}
        \includegraphics[width=\twopanelfigurewidth]{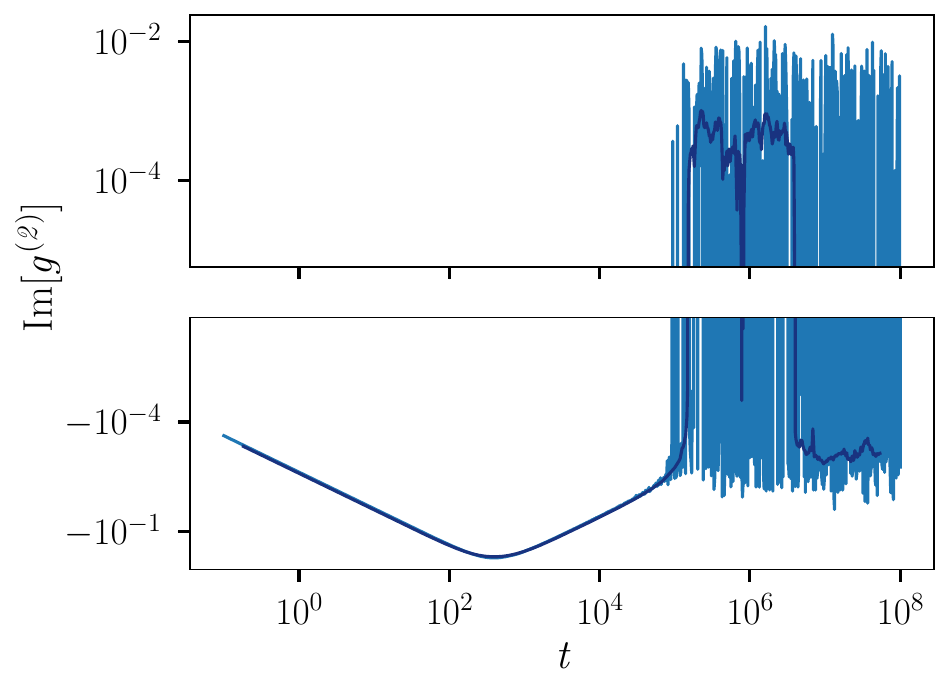}
    }\\
    \subfloat[de Sitter]
    {
        \includegraphics[width=\twopanelfigurewidth]{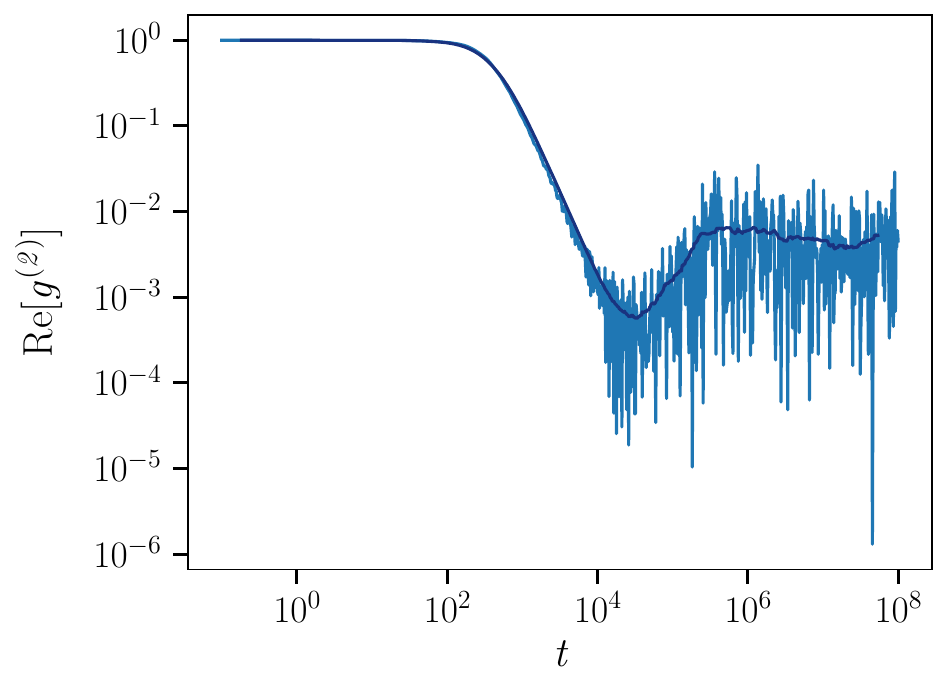}
        \includegraphics[width=\twopanelfigurewidth]{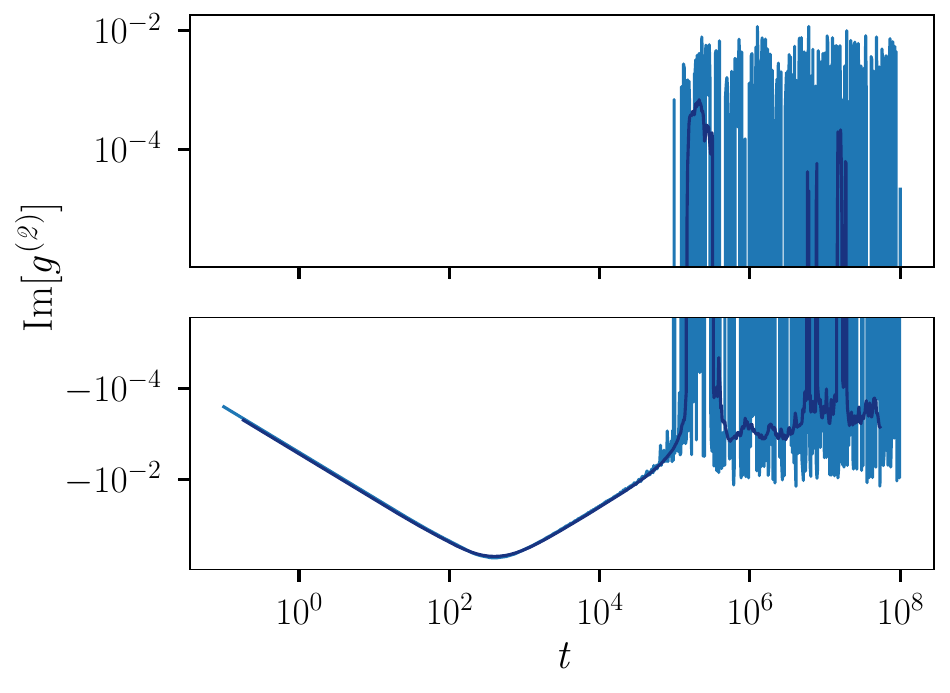}
    }
    \caption{The real and imaginary parts of the two-level generalized spectral form factor for the BTZ and de Sitter normal modes, ${\beta = 0}$, ${m = 0}$, ${J\cut = 200}$, ${n\cut = 1}$. With a darker line, the moving averaged data is shown. It can be noted that the oscillations of the imaginary part are not averaged into a plateau.}
    \label{fig:2lvlGSFFm0}
\end{figure}

In \figref{fig:2lvlGSFFm0}, we present the gSFF for massless scalar field on BTZ and de~Sitter background. The real part of the gSFF shows a similar behavior as the SFF, with a dip-ramp-plateau structure, which confirms a consistency of the modified quantity.

As opposed to the standard definition of the SFF, the gSFF has an imaginary part, which is shown in \figref{fig:2lvlGSFFm0} on the right. The imaginary part is negative when the real part dips while at large times, it oscillates rapidly around zero. Moving averaging of the oscillating part does not lead to a constant value plateau.

The similar behavior of plots for the gSFF can be observed for random matrices~\cite{Wei:2024ujf}. When possible, the RMT calculations can be performed analytically using an explicit form of the joint probability density functions. Alternatively, numerical calculation can be performed by a straightforward averaging over a finite ensemble. It turns out that the real part exhibits the same behaviors as dip, ramp, and plateau. The imaginary part starts from negative values. Then, it reaches a peak where the imaginary part becomes positive. In the later stage, the analytic result decreases to zero, while the numerical result shows an oscillating behavior around zero and a symmetric distribution. The oscillations appear due to the finiteness of the sample. The decline to zero of the imaginary part of the gSFF differs significantly from the late-time plateau of the real part. Therefore, the imaginary part coincides with that for numerical calculations on a finite ensemble of random matrices.

\begin{figure}[t!]\centering
    \subfloat[$n\cut = 1$]
    {
        \includegraphics[width=\twopanelfigurewidth]{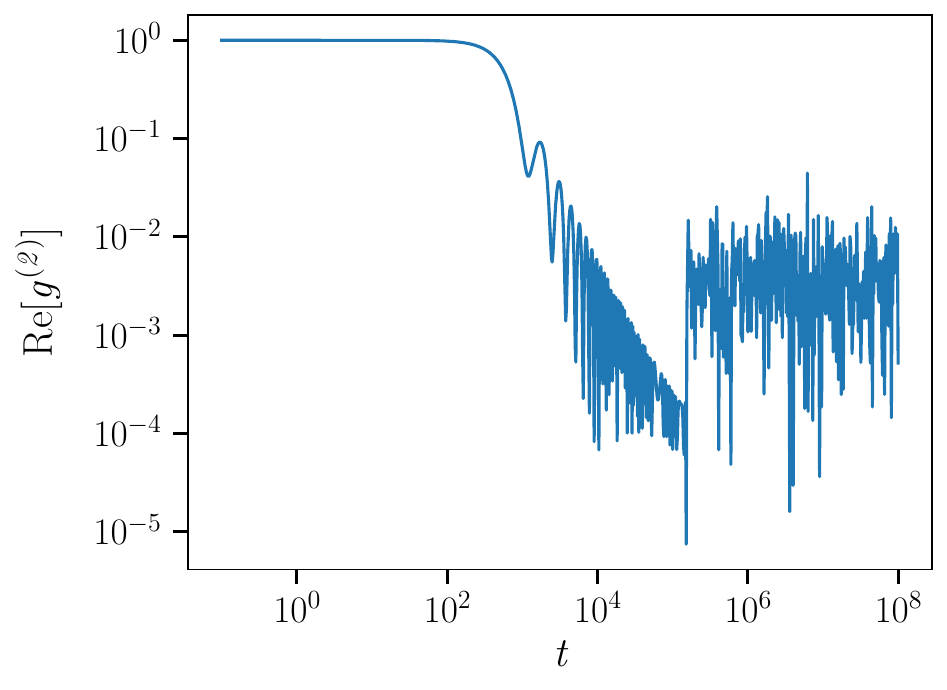}
        \includegraphics[width=\twopanelfigurewidth]{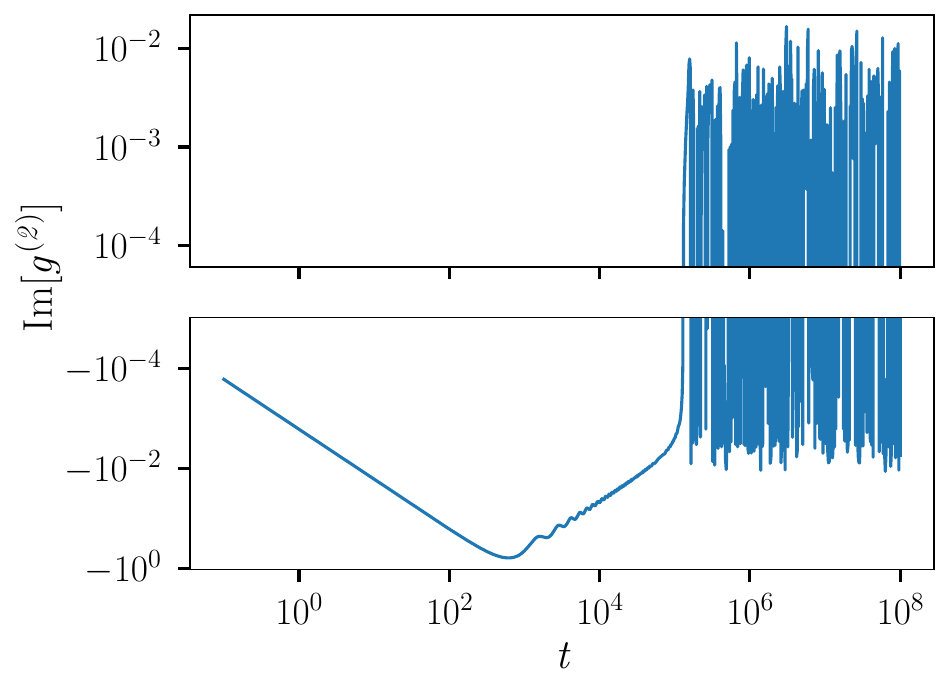}
    }\\
    \subfloat[$n\cut = 30$]
    {
        \includegraphics[width=\twopanelfigurewidth]{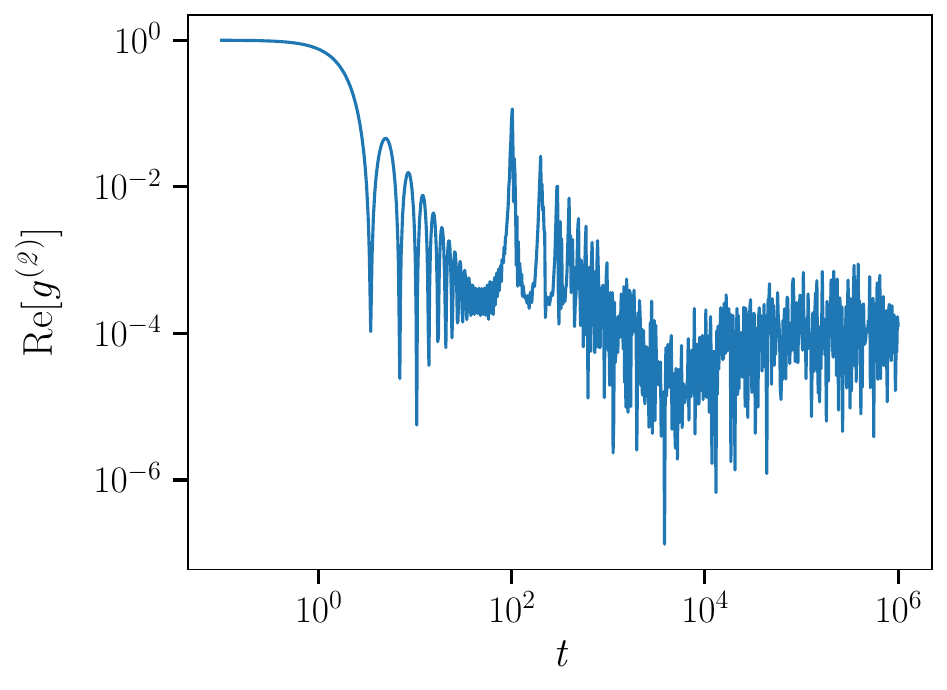}
        \includegraphics[width=\twopanelfigurewidth]{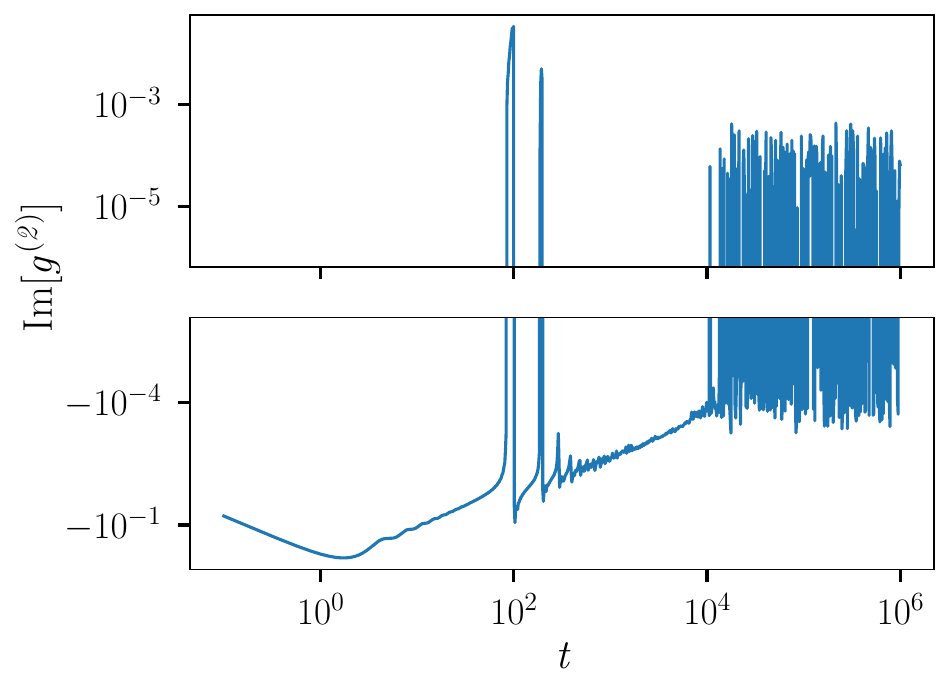}
    }
    \caption{The two-level gSFF for the massive scalar field on the de Sitter background, ${\beta = 0}, {m = 30}, {J\cut = 200}$, with different values of~$n\cut$.}
    \label{fig:2lvlGSFFm30dS}
\end{figure}

The effect of a non-zero mass is also consistent with the ordinary SFF. Specifically, for ${n\cut = 1}$ the ramp part vanishes, but as $n\cut$ increases, the dip-ramp-plateau structure is recovered, see \figref{fig:2lvlGSFFm30dS}.

%----------------------%
\subsubsection{Absence of the ramp in integrable models}

Let us consider the harmonic oscillator and the rectangular billiard as the two simplest examples of integrable systems. The corresponding spectra are given by
\be\label{eq:spectrumOscBill}
    \begin{aligned}
        &\text{Harmonic oscillator:} \quad &E_n = &\,\alpha\left(n + \frac{1}{2}\right),\\
        &\text{Rectangular billiard:} \quad &E_{n_1,\,n_2} = &\,\alpha_1 n_1^2 + \alpha_2 n_2^2.
    \end{aligned}
\ee

It is possible to obtain an analytic expression of the two-level gSFF for the harmonic oscillator due to the linearity of the spectrum. It has the following form:
\be
    \begin{aligned}
        \re[g^{(2)}(t)] &= \frac{1}{(n_\text{max} + 1)^2\sin^2\left(\frac{\alpha t}{4}\right)}\sin^2\left(\frac{n_\text{max} + 1}{4}\alpha t\right), \\
        \im[g^{(2)}(t)] &= -\frac{1}{2(n_\text{max} + 1)^2\sin^2\left(\frac{\alpha t}{4}\right)}\left[(n_\text{max} + 1)\sin\frac{\alpha t}{2} - \sin\left(\frac{n_\text{max} + 1}{2}\alpha t\right)\right],
    \end{aligned}
\ee
where $n_\text{max}$ is the spectrum cutoff.

\begin{figure}[t]\centering
    \subfloat[harmonic oscillator]
    {
        \includegraphics[width=\twopanelfigurewidth]{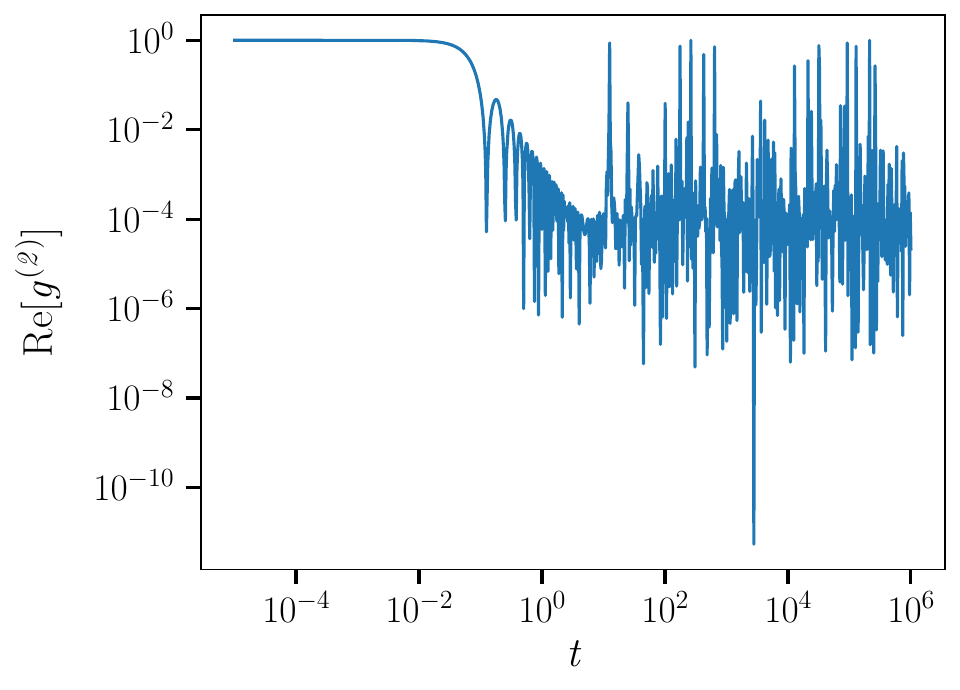}
        \includegraphics[width=\twopanelfigurewidth]{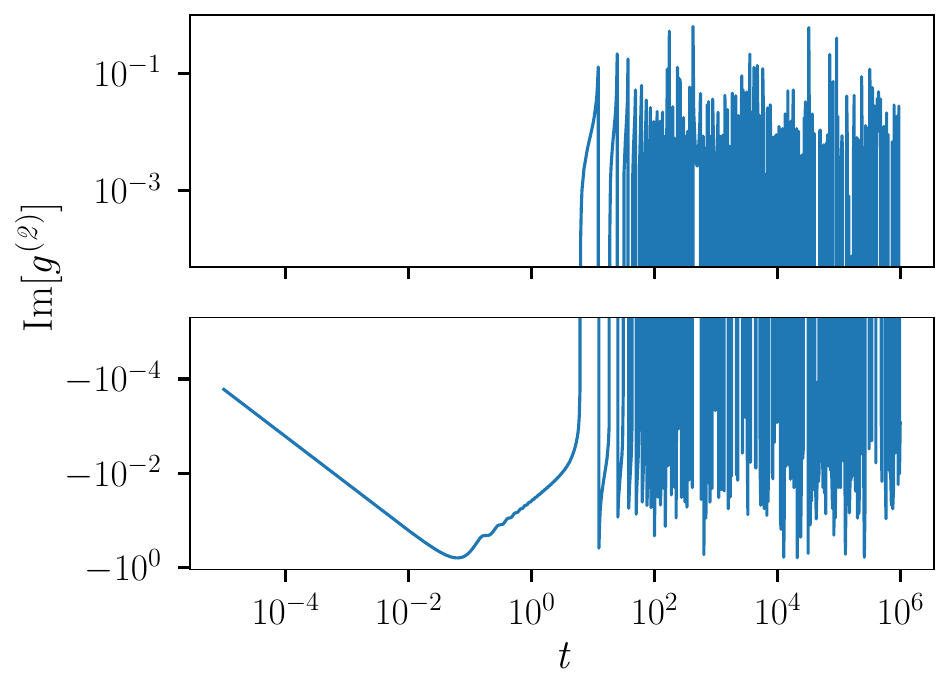}
    }
    \\
    \subfloat[rectangular billiard]
    {
        \includegraphics[width=\twopanelfigurewidth]{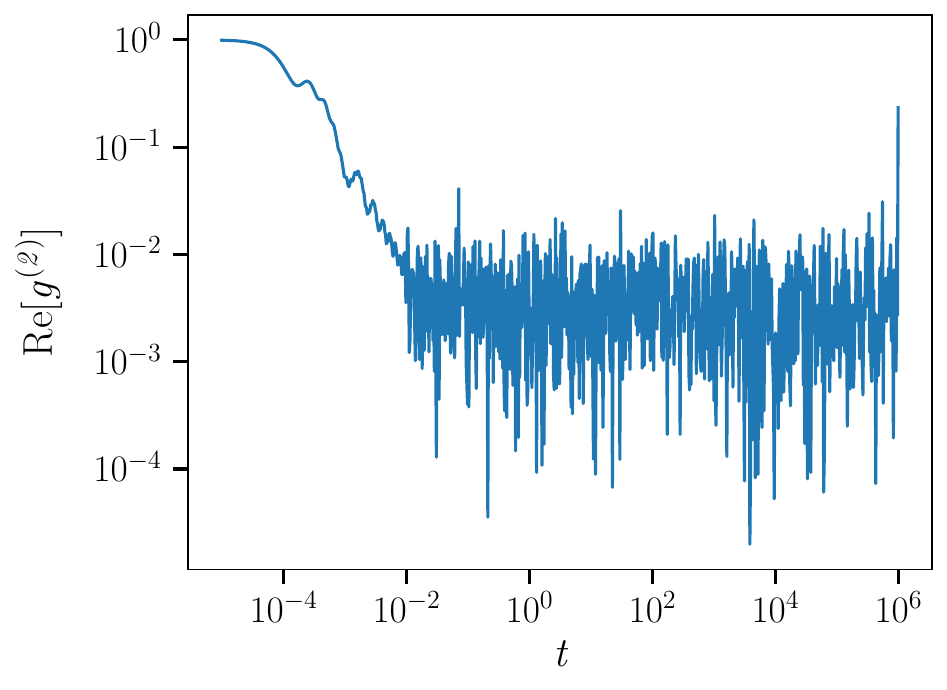}
        \includegraphics[width=\twopanelfigurewidth]{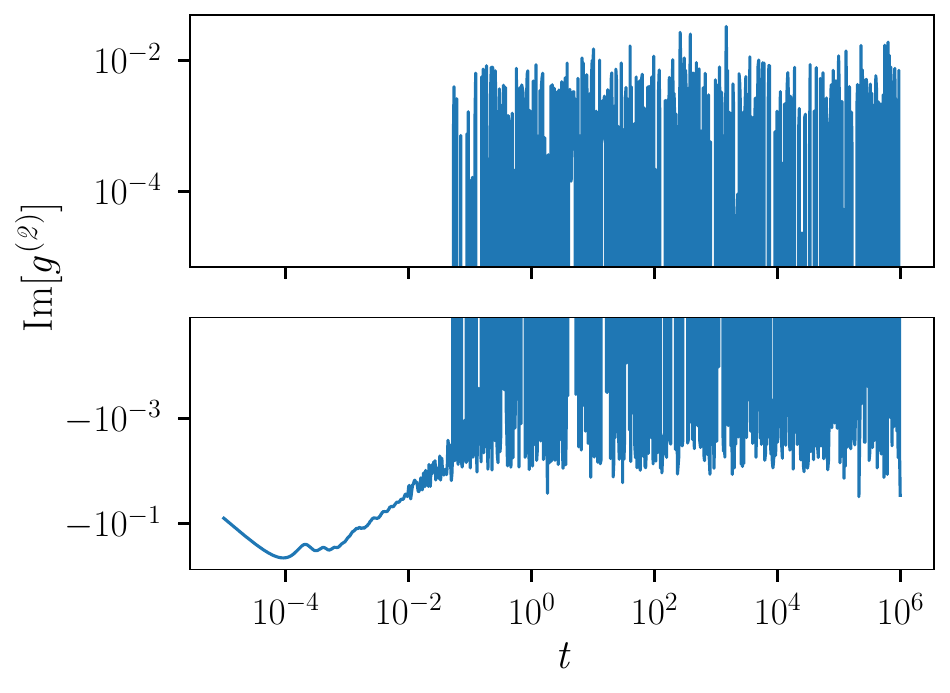}
    }
    \caption{The two-level gSFF for the simplest regular models: a) harmonic oscillator, the parameters are: ${n_\text{max} = 100}$ and ${\alpha = 1}$; b) rectangular billiard, the parameters are: $n_{1, \text{\,max}} = 100$, $n_{2, \text{\,max}} = 100$, $\alpha_1 = 1$ and $\alpha_2 = 2\pi$. The irrational ratio of $\alpha_{1,\,2}$ is chosen in order to ensure that the level spacing distribution is given by Poisson statistics~\cite{PhysRevLett.54.1350}.}
    \label{fig:2lvlReg}
\end{figure}

In \figref{fig:2lvlReg}, we have plotted the real and imaginary parts of the two-level gSFF for the harmonic oscillator and for the rectangular billiard, respectively. It is apparent that there is no dip-ramp-plateau structure on the plots of the real parts. This confirms that without additional averaging over the ensemble, the ramp does not exhibit in integrable models, in contrast to the BTZ and de Sitter scalar field normal modes. However, the imaginary part has a similar structure to that in models with a stretched horizon: there is a minimum of negative values, followed by rapid oscillations around zero.

%-------------------------%
\subsection{Three-level generalized spectral form factor}
\subsubsection{BTZ black hole and de Sitter}

\begin{figure}[t]\centering
     \subfloat[BTZ, $m = 0$]
     {
        \includegraphics[width=\twopanelfigurewidth]{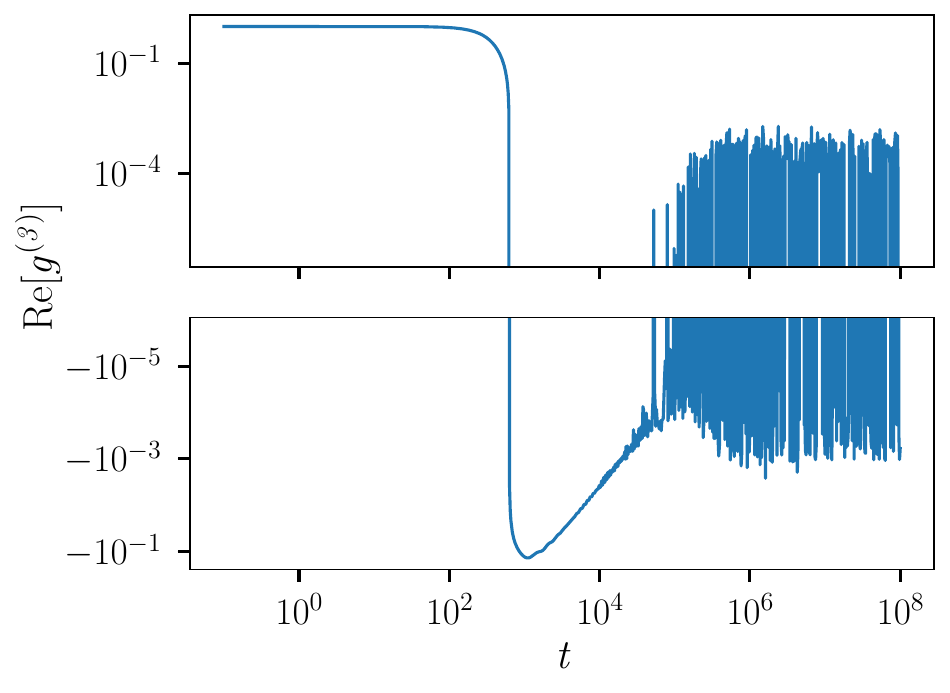}
        \includegraphics[width=\twopanelfigurewidth]{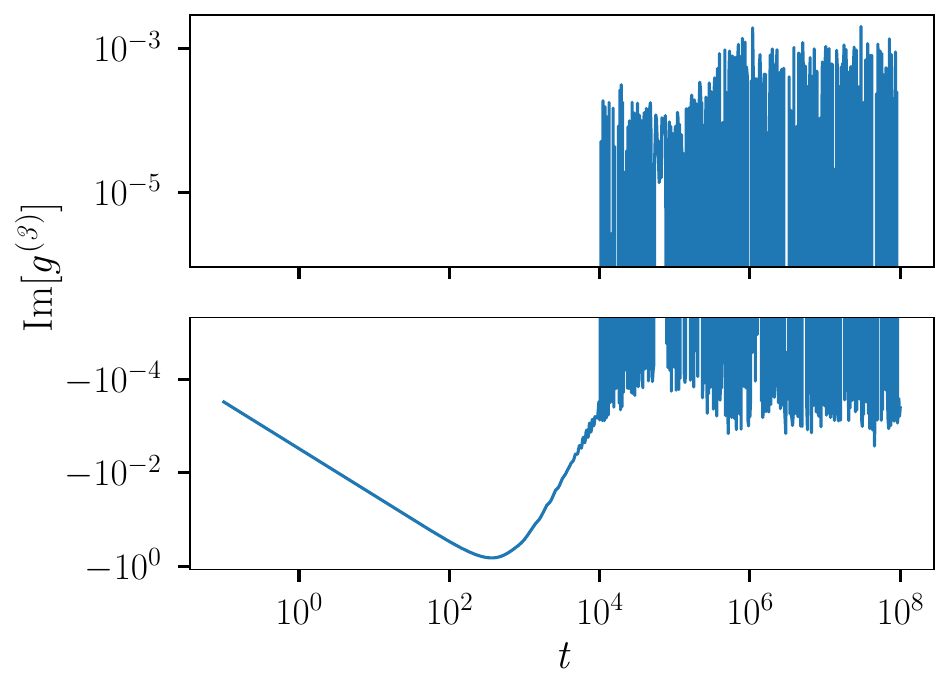}
    }\\
    \subfloat[de Sitter, $m = 0$]
    {
        \includegraphics[width=\twopanelfigurewidth]{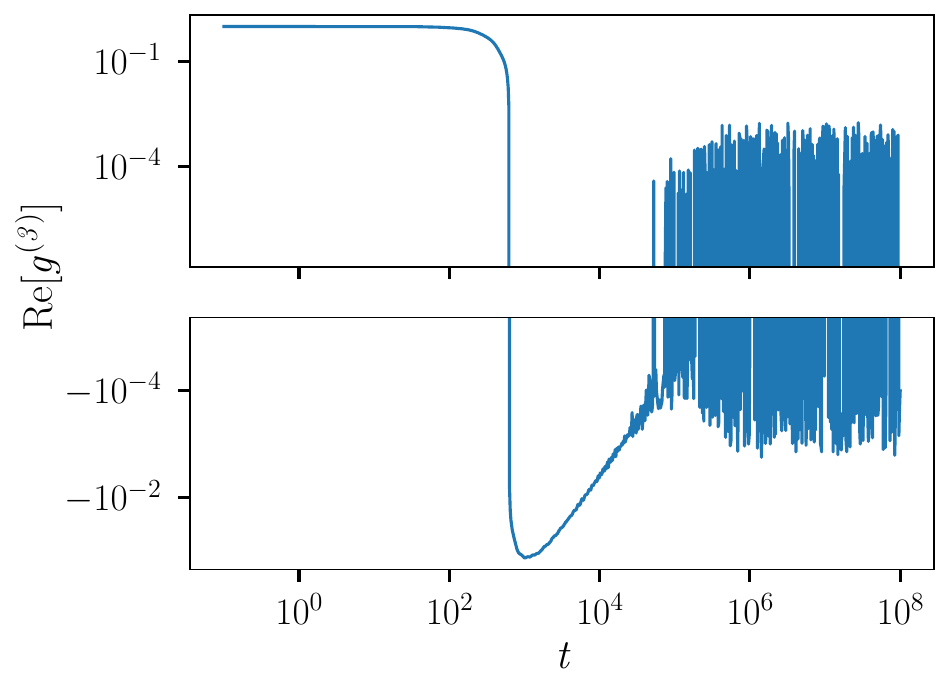}
        \includegraphics[width=\twopanelfigurewidth]{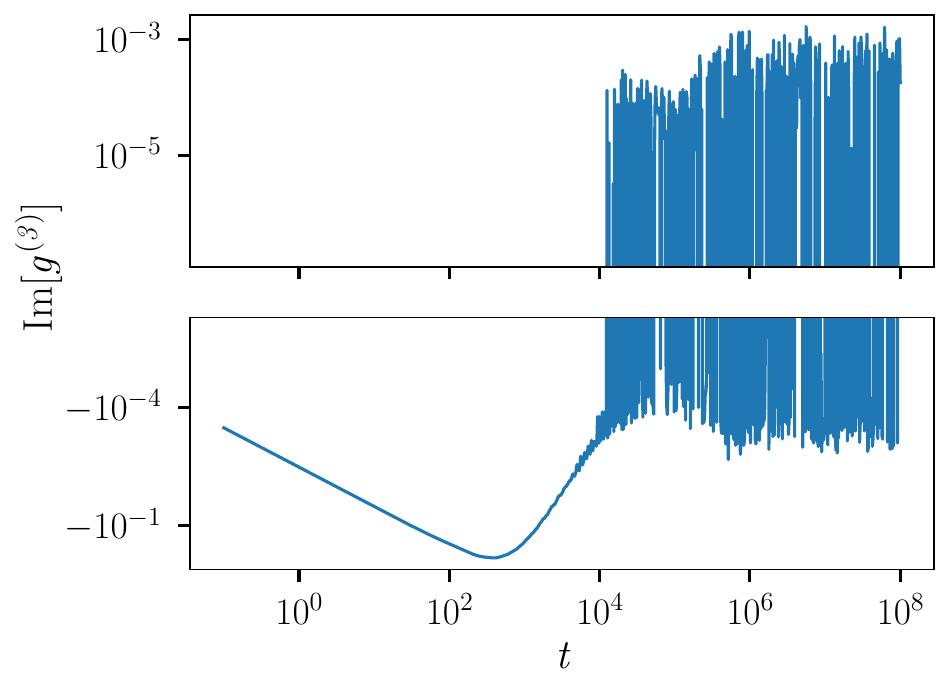}
    }
    \caption{The real and imaginary part of the three-level gSFF for massless scalar field on the BTZ and de Sitter backgrounds, ${\beta = 0}$, ${J\cut = 200}$, ${n\cut = 1}$.}
    \label{fig:3lvlN1}
\end{figure}

\begin{figure}[t]\centering
    \subfloat[$m = 0$]
    {
        \includegraphics[width=\twopanelfigurewidth]{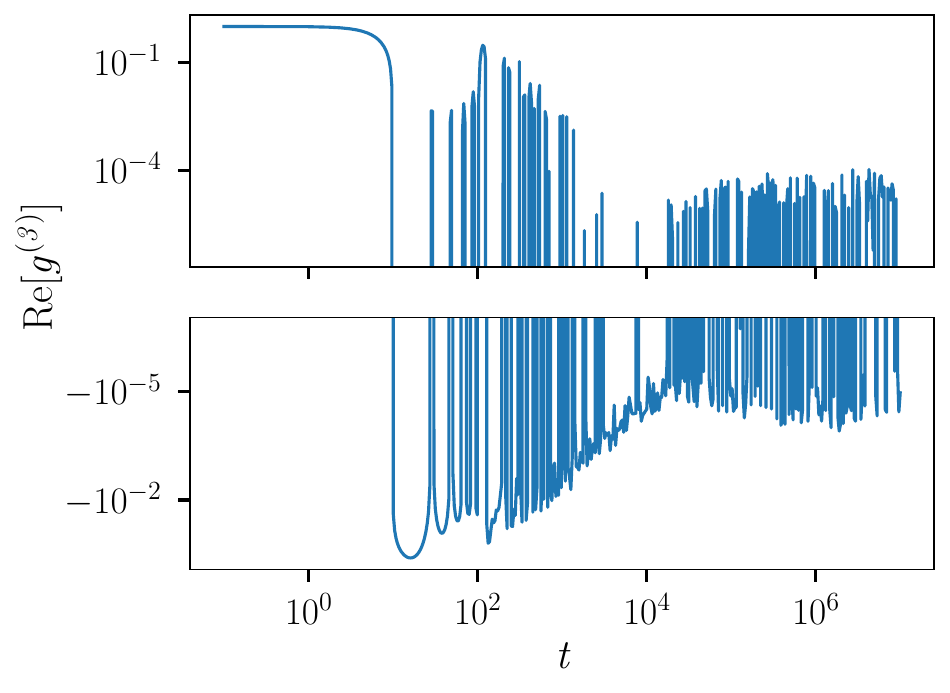}
        \includegraphics[width=\twopanelfigurewidth]{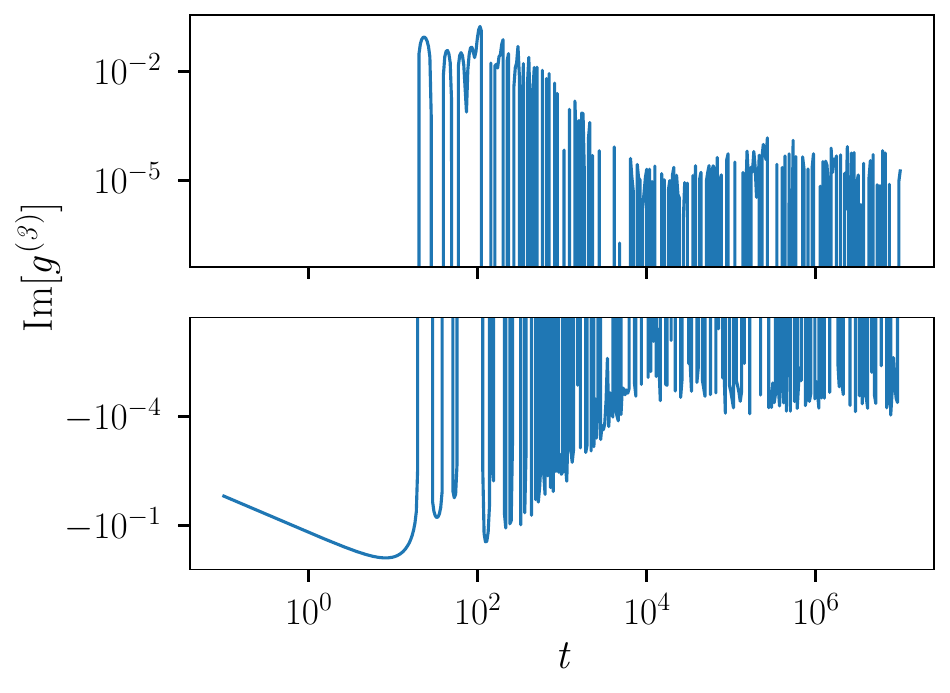}
    }\\
    \subfloat[$m = 60$]
    {
        \includegraphics[width=\twopanelfigurewidth]{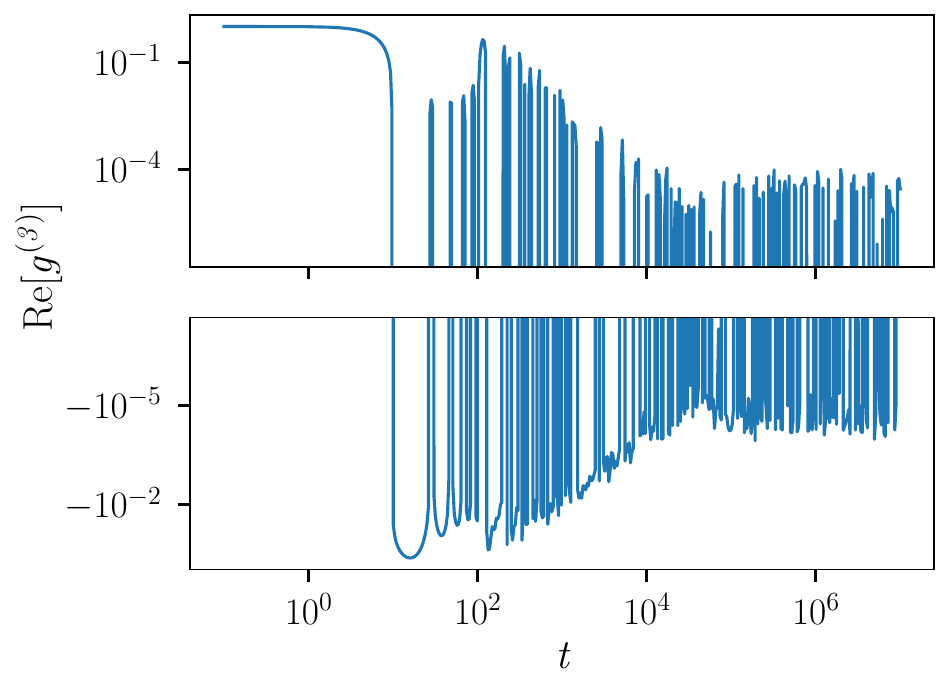}
        \includegraphics[width=\twopanelfigurewidth]{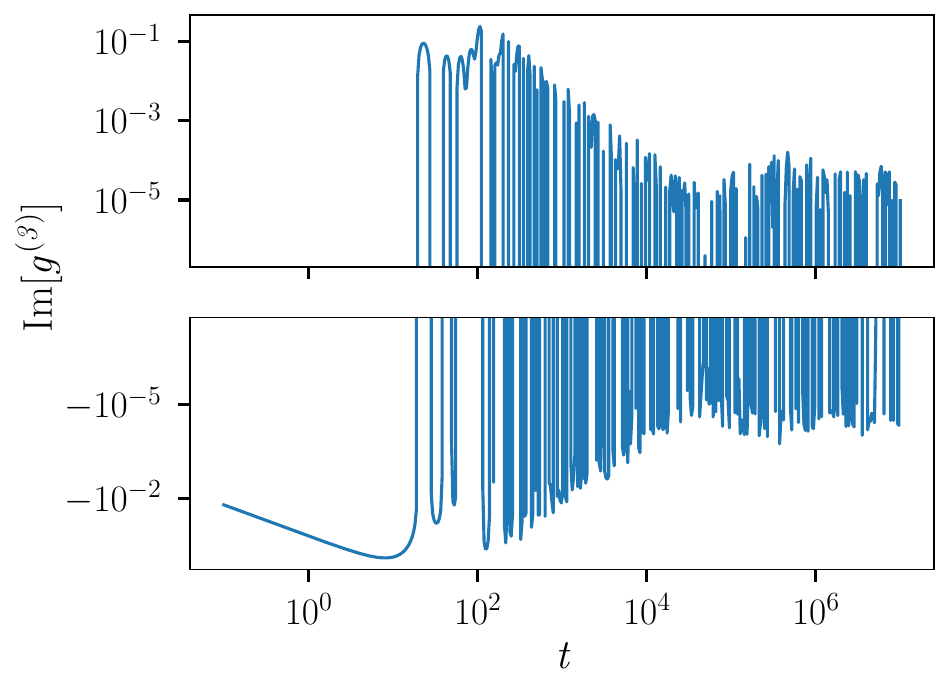}
    }
    \caption{The real and imaginary parts of the three-level gSFF for massless and massive (${m = 60}$) scalar field on the BTZ background, ${\beta = 0}$, ${J\cut = 200}$, ${n\cut = 6}$.}
    \label{fig:3lvlN6}
\end{figure}

The results for the real part of the three-level gSFF for the BTZ and de Sitter backgrounds are shown in Figures~\ref{fig:3lvlN1} and~\ref{fig:3lvlN6} on the left. As in the cases of the standard definition of the SFF and the two-level gSFF, the obtained structure is highly similar for the BTZ and de Sitter backgrounds.

The time dependence of the real part of the gSFF can be divided into three time periods. For small times, the real part of the three-level gSFF is a constant that equals unity. Then, there is a sharp drop into the negative region. In the third region, it oscillates around zero, which differs significantly from the behavior of the real part for random matrices described in~\cite{Wei:2024ujf}. Indeed, numerical calculations of the real part for random matrices show that, at large times, the real part becomes positive again and reaches a plateau. This means that even if the sample is finite, there are no oscillations. The plateau value depends on the dimension of the matrices as~$1/N^2$ and the real part vanishes at later times in the limit ${N \to \infty}$.

In Figures~\ref{fig:3lvlN1} and~\ref{fig:3lvlN6} on the right, we present the results for the imaginary part of the three-level gSFF, which shows a similar behavior to that of the two-level gSFF. At early stage, the imaginary part is negative, and then oscillates around zero. For ${n_\text{cut} > 1}$ (\figref{fig:3lvlN6} on the right), the similarity with the structure arising in numerical calculations in the theory of random matrices is particularly notable. The behavior of the imaginary part exhibits a common dip followed by an oscillatory decay and then oscillates around zero. However, in the theory of random matrices, these oscillations at large times are associated with the finite size~$N$ of the matrices. The larger the~$N$, the smaller the amplitude of these oscillations.

%------------------------------%
\subsubsection{Typical integrable models}

\begin{figure}[t]\centering
    \subfloat[harmonic oscillator]
    {
        \includegraphics[width=\twopanelfigurewidth]{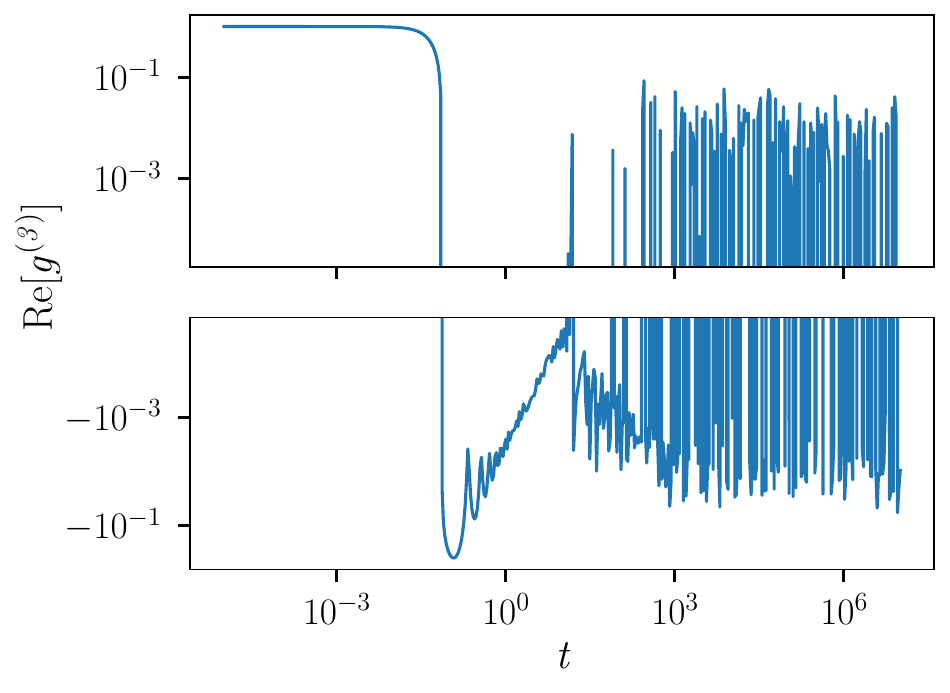}
        \includegraphics[width=\twopanelfigurewidth]{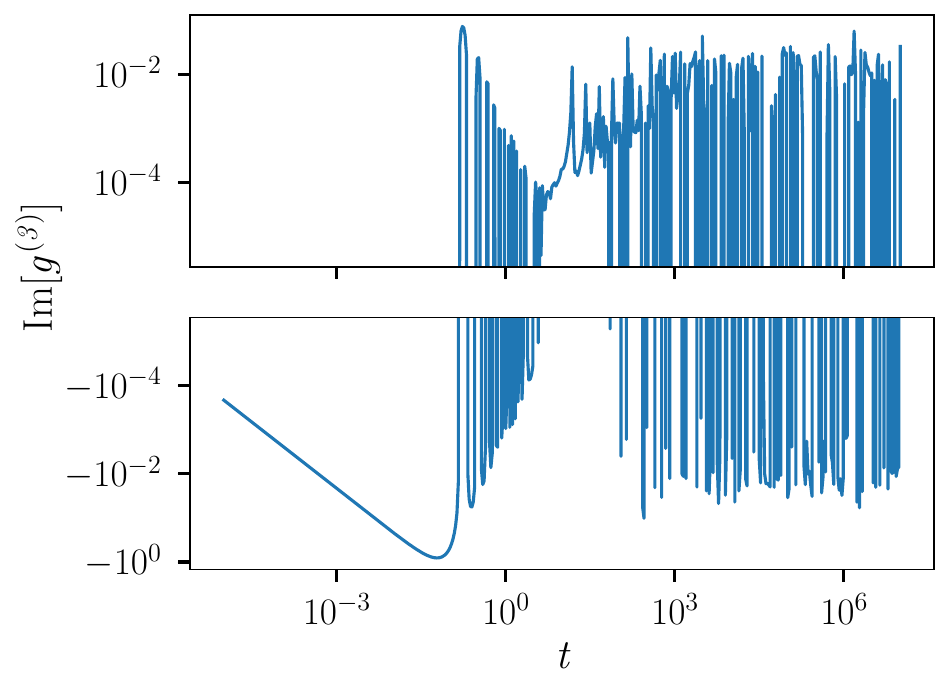}
    }\\
    \subfloat[rectangular billiard]
    {
        \includegraphics[width=\twopanelfigurewidth]{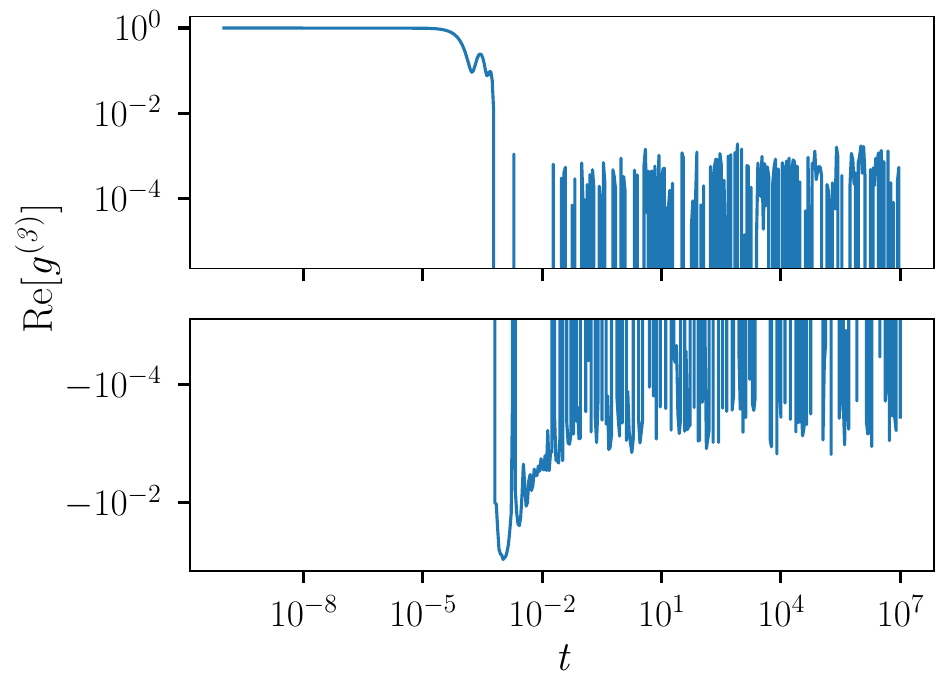}
        \includegraphics[width=\twopanelfigurewidth]{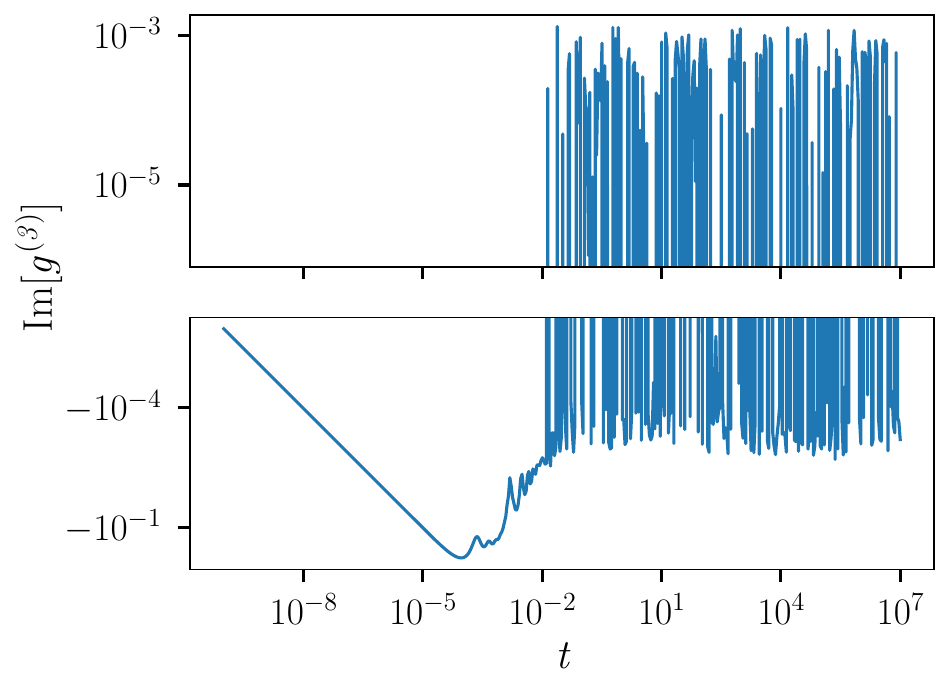}
    }
    \caption{The three-level gSFF for the simplest regular models: a) harmonic oscillator, the parameters are: ${n_\text{max} = 100}$ and ${\alpha = 1}$; b) rectangular billiard, the parameters are: $n_{1, \text{\,max}} = 100$, $n_{2, \text{\,max}} = 100$, $\alpha_1 = 1$ and $\alpha_2 = 2\pi$.}
    \label{fig:3lvlReg}
\end{figure}

In Figure~\ref{fig:3lvlReg}, we have plotted the real and imaginary parts of the three-level gSFF for the harmonic oscillator and for the rectangular billiard with corresponding spectra~\eqref{eq:spectrumOscBill}. In contrast to the models with a stretched horizon, there is noticeable non-trivial behavior for relatively small values of~$t$. There are differences in the early-time behavior of the gSFF for the oscillator compared to the BTZ and de Sitter normal modes for ${n\cut = 1}$, as well as in the behavior of the billiard compared to BTZ and de Sitter normal modes for ${n\cut = 6}$. At a later stage, rapid fluctuations around a value close to zero are also observed.

%--------------------------%
\section{Conclusions}

Let us summarize our findings. In this paper, we have studied generalizations of the spectral form factor for the normal modes of the massive scalar field in the BTZ black hole and de Sitter spacetimes. We have considered the ordinary SFF and the two- and three-level gSFF. We found that, despite the fact that these backgrounds correspond to opposite signs of the constant curvature and cosmological constant, the scalar field normal modes demonstrate a very similar form factor structure. In this way, we see that the specific SFF behaviour is characteristic of horizon, and is not determined by the negative curvature of the BTZ black hole.

The distinctive feature of the ordinary SFF for the massless field, namely, the dip-ramp-plateau shape, found in~\cite{Das:2022evy} for the BTZ black hole, appears also for de Sitter static patch as does the non-RMT behavior of the level spacing distribution. However, for the relatively small number of massive field normal modes on both backgrounds, the ramp part is almost absent, and the SFF abruptly reaches the plateau after the dip. Only a significant increase of the cutoff parameters of the spectrum helps to recover the DRP structure.

The structure of the two-level gSFF is consistent with the ordinary SFF and is analogous to that of the theory of random matrices, while the behavior of the three-level gSFF has a feature that distinguishes it from the  RMT result. This difference is manifested as the rapid fluctuations in the real and imaginary components at later times while at early times, the structure of the form factor is similar to that of random matrices. The same late-time fluctuations are intrinsic to simple integrable models whose form factor, however, differs from that of normal modes at earlier times.

\skipline

We can conclude that both the BTZ black hole and the de Sitter brick-wall models exhibit the signs of quantum chaos at the level of two-level SFF. The generalized form factor, being a more delicate characteristic, leads to the open question of whether this behavior is a manifestation of pseudo-chaos, when the RMT-like shape of chaotic measures appears for integrable models, or an indication that the ordinary SFF as a signature of chaos is not universal. The observed for the shape of the three-level gSFF differences make more preferable the second version because this, as well as the non-Gaussian level spacing distribution, is what distinguishes chaotic signatures in curved spacetime with a horizon from that in the theory of random matrices. Our conclusion concerns only models without disorder-averaged random boundary conditions, because the presence of such condition allows to successfully reproduce level-spacing of the random matrix ensembles.

%--------------------------%
\acknowledgments

We would like to thank Mihailo \v{C}ubrovi\'{c} for the thorough discussion and Aleksandr Belokon and Timofei Rusalev for valuable comments on the manuscript. This work was supported by the Russian Science Foundation under grant No.~24-72-10061,
\href{https://rscf.ru/en/project/24-72-10061/}{https://rscf.ru/en/project/24-72-10061/}.

%--------------------------%
\clearpage
\bibliographystyle{JHEP}
\bibliography{biblio}

\providecommand{\href}[2]{#2}\begingroup\raggedright\begin{thebibliography}{10}

\bibitem{Bousso:2022ntt}
R.~Bousso, X.~Dong, N.~Engelhardt, T.~Faulkner, T.~Hartman, S.H.~Shenker et~al., \emph{{Snowmass White Paper: Quantum Aspects of Black Holes and the Emergence of Spacetime}},  \href{https://arxiv.org/abs/2201.03096}{{\ttfamily 2201.03096}}.

\bibitem{Sekino:2008he}
Y.~Sekino and L.~Susskind, \emph{{Fast Scramblers}}, \href{https://doi.org/10.1088/1126-6708/2008/10/065}{\emph{JHEP} {\bfseries 10} (2008) 065} [\href{https://arxiv.org/abs/0808.2096}{{\ttfamily 0808.2096}}].

\bibitem{Shenker:2013pqa}
S.H.~Shenker and D.~Stanford, \emph{{Black holes and the butterfly effect}}, \href{https://doi.org/10.1007/JHEP03(2014)067}{\emph{JHEP} {\bfseries 03} (2014) 067} [\href{https://arxiv.org/abs/1306.0622}{{\ttfamily 1306.0622}}].

\bibitem{Maldacena:2015waa}
J.~Maldacena, S.H.~Shenker and D.~Stanford, \emph{{A bound on chaos}}, \href{https://doi.org/10.1007/JHEP08(2016)106}{\emph{JHEP} {\bfseries 08} (2016) 106} [\href{https://arxiv.org/abs/1503.01409}{{\ttfamily 1503.01409}}].

\bibitem{Haake:2010fgh}
F.~Haake, \emph{{Quantum Signatures of Chaos}}, Springer Series in Synergetics, Springer, Berlin (2010), \href{https://doi.org/10.1007/978-3-642-05428-0}{10.1007/978-3-642-05428-0}.

\bibitem{Bohigas84}
O.~Bohigas, M.J.~Giannoni and C.~Schmit, \emph{Characterization of chaotic quantum spectra and universality of level fluctuation laws}, \href{https://doi.org/10.1103/PhysRevLett.52.1}{\emph{Phys. Rev. Lett.} {\bfseries 52} (1984) 1}.

\bibitem{Rosenhaus:2020tmv}
V.~Rosenhaus, \emph{{Chaos in the Quantum Field Theory S-Matrix}}, \href{https://doi.org/10.1103/PhysRevLett.127.021601}{\emph{Phys. Rev. Lett.} {\bfseries 127} (2021) 021601} [\href{https://arxiv.org/abs/2003.07381}{{\ttfamily 2003.07381}}].

\bibitem{Savic:2024ock}
N.~Savi\'c and M.~\v{C}ubrovi\'c, \emph{{Weak chaos and mixed dynamics in the string S-matrix}}, \href{https://doi.org/10.1007/JHEP03(2024)101}{\emph{JHEP} {\bfseries 03} (2024) 101} [\href{https://arxiv.org/abs/2401.02211}{{\ttfamily 2401.02211}}].

\bibitem{Bianchi:2022mhs}
M.~Bianchi, M.~Firrotta, J.~Sonnenschein and D.~Weissman, \emph{{Measure for Chaotic Scattering Amplitudes}}, \href{https://doi.org/10.1103/PhysRevLett.129.261601}{\emph{Phys. Rev. Lett.} {\bfseries 129} (2022) 261601} [\href{https://arxiv.org/abs/2207.13112}{{\ttfamily 2207.13112}}].

\bibitem{Bianchi:2023uby}
M.~Bianchi, M.~Firrotta, J.~Sonnenschein and D.~Weissman, \emph{{Measuring chaos in string scattering processes}}, \href{https://doi.org/10.1103/PhysRevD.108.066006}{\emph{Phys. Rev. D} {\bfseries 108} (2023) 066006} [\href{https://arxiv.org/abs/2303.17233}{{\ttfamily 2303.17233}}].

\bibitem{berry2005riemann}
M.V.~Berry, \emph{Riemann's zeta function: A model for quantum chaos?},  in \emph{Quantum Chaos and Statistical Nuclear Physics: Proceedings of the 2nd International Conference on Quantum Chaos and the 4th International Colloquium on Statistical Nuclear Physics, Held at Cuernavaca, M{\'e}xico, January 6--10, 1986}, pp.~1--17, Springer, 2005.

\bibitem{odlyzko1987distribution}
A.M.~Odlyzko, \emph{On the distribution of spacings between zeros of the zeta function}, {\emph{Mathematics of Computation} {\bfseries 48} (1987) 273}.

\bibitem{Liu:2018hlr}
J.~Liu, \emph{{Spectral form factors and late time quantum chaos}}, \href{https://doi.org/10.1103/PhysRevD.98.086026}{\emph{Phys. Rev. D} {\bfseries 98} (2018) 086026} [\href{https://arxiv.org/abs/1806.05316}{{\ttfamily 1806.05316}}].

\bibitem{Cotler:2016fpe}
J.S.~Cotler, G.~Gur-Ari, M.~Hanada, J.~Polchinski, P.~Saad, S.H.~Shenker et~al., \emph{{Black Holes and Random Matrices}}, \href{https://doi.org/10.1007/JHEP05(2017)118}{\emph{JHEP} {\bfseries 05} (2017) 118} [\href{https://arxiv.org/abs/1611.04650}{{\ttfamily 1611.04650}}].

\bibitem{Das:2022evy}
S.~Das, C.~Krishnan, A.P.~Kumar and A.~Kundu, \emph{{Synthetic fuzzballs: a linear ramp from black hole normal modes}}, \href{https://doi.org/10.1007/JHEP01(2023)153}{\emph{JHEP} {\bfseries 01} (2023) 153} [\href{https://arxiv.org/abs/2208.14744}{{\ttfamily 2208.14744}}].

\bibitem{Das:2023ulz}
S.~Das, S.K.~Garg, C.~Krishnan and A.~Kundu, \emph{{Fuzzballs and random matrices}}, \href{https://doi.org/10.1007/JHEP10(2023)031}{\emph{JHEP} {\bfseries 10} (2023) 031} [\href{https://arxiv.org/abs/2301.11780}{{\ttfamily 2301.11780}}].

\bibitem{Das:2023xjr}
S.~Das and A.~Kundu, \emph{{Brickwall in rotating BTZ: a dip-ramp-plateau story}}, \href{https://doi.org/10.1007/JHEP02(2024)049}{\emph{JHEP} {\bfseries 02} (2024) 049} [\href{https://arxiv.org/abs/2310.06438}{{\ttfamily 2310.06438}}].

\bibitem{Krishnan:2023jqn}
C.~Krishnan and P.S.~Pathak, \emph{{Normal modes of the stretched horizon: a bulk mechanism for black hole microstate level spacing}}, \href{https://doi.org/10.1007/JHEP03(2024)162}{\emph{JHEP} {\bfseries 03} (2024) 162} [\href{https://arxiv.org/abs/2312.14109}{{\ttfamily 2312.14109}}].

\bibitem{Banerjee:2024dpl}
S.~Banerjee, S.~Das, M.~Dorband and A.~Kundu, \emph{{Brickwall, normal modes, and emerging thermality}}, \href{https://doi.org/10.1103/PhysRevD.109.126020}{\emph{Phys. Rev. D} {\bfseries 109} (2024) 126020} [\href{https://arxiv.org/abs/2401.01417}{{\ttfamily 2401.01417}}].

\bibitem{Jeong:2024oao}
H.-S.~Jeong, A.~Kundu and J.F.~Pedraza, \emph{{Brickwall One-Loop Determinant: Spectral Statistics \& Krylov Complexity}},  \href{https://arxiv.org/abs/2412.12301}{{\ttfamily 2412.12301}}.

\bibitem{Banerjee:2024ivh}
S.~Banerjee, S.~Das, A.~Kundu and M.~Sittinger, \emph{{Blackish Holes}},  \href{https://arxiv.org/abs/2411.09500}{{\ttfamily 2411.09500}}.

\bibitem{Wei:2024ujf}
Z.~Wei, C.~Tan and R.~Zhang, \emph{{Generalized spectral form factor in random matrix theory}}, \href{https://doi.org/10.1103/PhysRevE.109.064208}{\emph{Phys. Rev. E} {\bfseries 109} (2024) 064208} [\href{https://arxiv.org/abs/2401.02119}{{\ttfamily 2401.02119}}].

\bibitem{Ichinose:1994rg}
I.~Ichinose and Y.~Satoh, \emph{{Entropies of scalar fields on three-dimensional black holes}}, \href{https://doi.org/10.1016/0550-3213(95)00197-Z}{\emph{Nucl. Phys. B} {\bfseries 447} (1995) 340} [\href{https://arxiv.org/abs/hep-th/9412144}{{\ttfamily hep-th/9412144}}].

\bibitem{tHooft:1984kcu}
G.~'t~Hooft, \emph{{On the Quantum Structure of a Black Hole}}, \href{https://doi.org/10.1016/0550-3213(85)90418-3}{\emph{Nucl. Phys. B} {\bfseries 256} (1985) 727}.

\bibitem{Keski-Vakkuri:1998gmz}
E.~Keski-Vakkuri, \emph{{Bulk and boundary dynamics in BTZ black holes}}, \href{https://doi.org/10.1103/PhysRevD.59.104001}{\emph{Phys. Rev. D} {\bfseries 59} (1999) 104001} [\href{https://arxiv.org/abs/hep-th/9808037}{{\ttfamily hep-th/9808037}}].

\bibitem{AbramowitzStegun1965}
M.~Abramowitz and I.A.~Stegun, \emph{Handbook of mathematical functions with formulas, graphs, and mathematical tables}, {\emph{National Bureau of Standards Applied Mathematics Series} {\bfseries 55} (1965) 953}.

\bibitem{Abdalla:2002rm}
E.~Abdalla, B.~Wang, A.~Lima-Santos and W.G.~Qiu, \emph{{Support of dS / CFT correspondence from perturbations of three-dimensional space-time}}, \href{https://doi.org/10.1016/S0370-2693(02)02039-7}{\emph{Phys. Lett. B} {\bfseries 538} (2002) 435} [\href{https://arxiv.org/abs/hep-th/0204030}{{\ttfamily hep-th/0204030}}].

\bibitem{Oganesyan:2007wpd}
V.~Oganesyan and D.A.~Huse, \emph{{Localization of interacting fermions at high temperature}}, \href{https://doi.org/10.1103/PhysRevB.75.155111}{\emph{Phys. Rev. B} {\bfseries 75} (2007) 155111}.

\bibitem{Atas:2013gvn}
Y.Y.~Atas, E.~Bogomolny, O.~Giraud and G.~Roux, \emph{{Distribution of the Ratio of Consecutive Level Spacings in Random Matrix Ensembles}}, \href{https://doi.org/10.1103/PhysRevLett.110.084101}{\emph{Phys. Rev. Lett.} {\bfseries 110} (2013) 084101}.

\bibitem{PhysRevE.55.3886}
P.~Shukla, \emph{Higher order correlations in quantum chaotic spectra}, \href{https://doi.org/10.1103/PhysRevE.55.3886}{\emph{Phys. Rev. E} {\bfseries 55} (1997) 3886}.

\bibitem{PhysRevLett.54.1350}
G.~Casati, B.V.~Chirikov and I.~Guarneri, \emph{Energy-level statistics of integrable quantum systems}, \href{https://doi.org/10.1103/PhysRevLett.54.1350}{\emph{Phys. Rev. Lett.} {\bfseries 54} (1985) 1350}.

\end{thebibliography}\endgroup

\end{document}